\documentclass[11pt,a4paper]{article}            
 \pdfoutput=1           

\usepackage{a4}          

\usepackage{hyperref}    

\usepackage{amsmath}
\usepackage{amssymb}
\usepackage[usenames]{color} 
\usepackage{graphicx}
\usepackage{array}
\usepackage{hhline}
\usepackage{microtype}
\usepackage[bf]{caption}
\usepackage{color}
\usepackage[usenames,dvipsnames]{xcolor}
\usepackage{tcolorbox}
\usepackage{multirow}
 \usepackage{cancel,slashed}

\def\hybrid{\topmargin -20pt    \oddsidemargin 0pt
        \headheight 0pt \headsep 0pt
        \textwidth 6.25in       
        \textheight 9 in       
        \marginparwidth .875in
        \parskip 5pt plus 1pt 
          \jot = 1.5ex
   }
\hybrid
\numberwithin{equation}{section}
\numberwithin{table}{section}

\newcommand{\beq}{\begin{equation}}  \newcommand{\eeq}{\end{equation}}
\newcommand{\bal}{\begin{aligned}}   \newcommand{\eal}{\end{aligned}}
\newcommand{\bea}{\begin{eqnarray}}  \newcommand{\eea}{\end{eqnarray}}

\newcommand{\bmat}{\left(\begin{array}}
\newcommand{\emat}{\end{array}\right)}
\def\om{\omega}

\newcommand{\cO}{\mathcal{O}}

\newcommand{\cK}{\mathcal{K}}
\newcommand{\cN}{\mathcal{N}}

\newcommand{\cR}{\mathcal{R}}

\newcommand{\cV}{\mathcal{V}}

\def\oh{\frac{1}{2}}
\def\a{{\alpha}}
\def\b{{\beta}}
\def\d{{\delta}}

\def\th{{\theta}}

\def\Om{{\Omega}}
\def\om{{\omega}}

\def\p{{\partial}}

\def\IZ{\mathbb{Z}}

\def\re{\mbox{Re }}
\def\im{\mbox{Im }}

\newcommand{\be}{\begin{equation}}
\newcommand{\ee}{\end{equation}}

\begin{document}

\baselineskip=14pt
\parskip 5pt plus 1pt

\vspace*{-1.5cm}
\begin{flushright}    
  {\small
IFT-UAM/CSIC-20-51}
\end{flushright}

\vspace{2cm}
\begin{center}        
  {\Huge On supersymmetric AdS$_4$ orientifold vacua}
\end{center}

\vspace{0.5cm}
\begin{center}        
{\large  Fernando Marchesano,$^a$ Eran Palti,$^{b,c}$ Joan Quirant,$^a$ and Alessandro Tomasiello$^{d,e}$}
\end{center}

\vspace{0.15cm}
\begin{center}        
\emph{$^a$ \small{Instituto de F\'{\i}sica Te\'orica UAM-CSIC, Cantoblanco, 28049 Madrid, Spain}}\\[.3cm]
\emph{$^b$ \small{Max-Planck-Institut f\"ur Physik (Werner-Heisenberg-Institut), 80805 M\"unchen, Germany}}\\[.3cm]
\emph{$^c$ \small{Department of Physics, Ben-Gurion University of the Negev, Beer-Sheva 84105, Israel}}\\[.3cm]
\emph{$^d$ \small{Dipartimento di Fisica, Universit\`a di Milano–Bicocca, Piazza della Scienza 3, I-20126 Milano, Italy}\\[.3cm]
\emph{$^e$INFN, sezione di Milano–Bicocca, Piazza della Scienza 3, I-20126 Milano, Italy}} 
             \\[0.15cm]
 
\end{center}

\vspace{2cm}


\begin{abstract}
\noindent In this work we study ten-dimensional solutions to type IIA string theory of the form AdS$_4 \times X_6$ which contain orientifold planes and preserve ${\cal N}=1$ supersymmetry. In particular, we consider solutions which exhibit some key features of the four-dimensional DGKT proposal for compactifications on Calabi--Yau manifolds with fluxes, and in this sense may be considered their ten-dimensional uplifts. We focus on the supersymmetry equations and Bianchi identities, and find solutions to these that are valid at the two-derivative level and at first order in an expansion parameter which is related to the AdS cosmological constant. This family of solutions is such that the background metric is deformed from the Ricci-flat one to one exhibiting $SU(3) \times SU(3)$-structure, and dilaton gradients and warp factors are induced.  
\end{abstract}

\thispagestyle{empty}
\clearpage

\tableofcontents

\setcounter{page}{1}



\section{Introduction}

String theory is known to support many Anti-de Sitter (AdS) vacua, namely solutions of the form ${\rm AdS}_d \times {\cal M}_p$ where all fields are invariant under the AdS isometries. 
Strikingly, for the vast majority of AdS vacua the Kaluza--Klein (KK) scale is comparable to the scale of the cosmological constant: one often says that there is no ``scale separation''. This means that the solutions are not really $d$-dimensional in any physical sense: physics looks ten- or eleven-dimensional to a hypothetical observer. There have been many studies of the property of scale separation in string theory, see in particular \cite{Tsimpis:2012tu,Gautason:2015tig,Gautason:2018gln,Blumenhagen:2019vgj,Font:2019uva,Apruzzi:2019ecr}.

Recently, the feature of scale separation was revisited as part of the Swampland program \cite{Vafa:2005ui} (see \cite{Brennan:2017rbf,Palti:2019pca} for reviews). In particular, it formed part of work addressing general properties of AdS space in quantum gravity \cite{Lust:2019zwm}. There,  it was suggested that a Swampland condition could be that the value of the cosmological constant sets the mass scale of an infinite tower of states. The AdS Distance Conjecture (ADC) states that this mass scale $m$ is related to the cosmological constant as 
\be
m \sim \Lambda^{\alpha} \;,
\ee
with $\alpha \sim {\cal O}(1)$. The conjecture was motivated by examples in string theory, but also by the fact that the $\Lambda \rightarrow 0$ limit is infinite distance in the space of metrics. Further, a Strong version of this conjecture was also proposed which states that for supersymmetric vacua $\alpha = \frac12$. This stronger form would be satisfied in any AdS vacuum which has no separation of scales.\footnote{The absence of scale separation was actually shown for general classical supersymmetric AdS$_7$ vacua in \cite{Apruzzi:2019ecr}.} The interesting proposal of the ADC, and more specifically its strong version, motivates the work in this paper.

We are particularly interested in a proposal for a set of vacua with scale separation which was put forward long ago in \cite{DeWolfe:2005uu} (DGKT), see also \cite{Camara:2005dc}.
These are AdS$_4$ vacua in type IIA string theory, have ${\mathcal N}=1$ supersymmetry, and have O6-plane singularities; indeed O-planes are supposed to be necessary for scale separation \cite{Gautason:2015tig}. The vacua were constructed by considering compactifications of type IIA string theory on a Calabi--Yau manifold $X_6$ in the presence of background fluxes. This led to an effective supersymmetric four-dimensional theory with a superpotential and K\"ahler potential that admit an infinite family of supersymmetric AdS$_4$ vacua. The theory is constructed by first considering a compactification of type IIA string theory on a Calabi--Yau manifold, which gives an ${\cal N}=2$ supersymmetric four-dimensional supergravity theory. The effective theory is then acted on by a projection which takes it to an ${\cal N}=1$ theory, as studied generally in \cite{Grimm:2004ua}. This projection captures the introduction of orientifolds (O6-planes), that act as sources of negative tension. Finally, a superpotential is induced in the theory, which aims at capturing the effects of turning on background fluxes, and follows the general form in \cite{Grimm:2004ua}. The resulting vacuum is then proposed to capture some full ten-dimensional solution of string theory. 

The four-dimensional construction of DGKT suggests a possible candidate counter-example to the Strong ADC. This was already discussed in \cite{Lust:2019zwm} where it was argued that since there is no known ten-dimensional uplift of this vacuum its features are not established and therefore may not be trustable. The aim of this work is to take some initial steps towards improving our understanding of ten-dimensional solutions which are based on the DGKT proposal.\footnote{Another approach towards establishing their validity would be to construct a dual CFT, which would have the so far unrealised features of a parametric hierarchy between the central charge and the scaling dimensions of an infinite number of operators. An initial search for the dual CFT to \cite{DeWolfe:2005uu} was carried out in \cite{Aharony:2008wz}.} 

Over the years, several attempts have been made to lift the four-dimensional DGKT construction to a ten-dimensional solution. The main difficulty lies in the presence of the O-plane sources. There exist several AdS solutions with back-reacted O-planes (but without scale separation: see for example \cite{Apruzzi:2019ecr} for a discussion in AdS$_7$). But in this case, the most concrete examples proposed in \cite{DeWolfe:2005uu} involve \emph{intersecting} O-planes, whose back-reaction isn't even known in flat space. In \cite{Acharya:2006ne} it was proposed to simply smear the O-planes; with this trick, an uplift to ten dimensions can indeed be found. Other similar solutions were found with the same trick in \cite{Petrini:2013ika}.

Smearing O-planes is not physically sensible though, so the next step was to investigate whether a similar solution could be found, where the O-planes could be localised. In \cite{Saracco:2012wc}, a local solution was found as a candidate for the behaviour near the individual O6-planes, with a resolved singularity and a large-distance asymptotics to the smeared solution of \cite{Acharya:2006ne}. However, it was not clear whether it could be made global; this partially motivated scepticism about the solution \cite{McOrist:2012yc}. 

Our approach to looking for a solution is to utilise the supersymmetry equations. First, we restrict to the equations at the two-derivative level, so neglecting higher order $\alpha'$ corrections. The supersymmetry is related to the structure group of the manifold. The Ricci-flat metric on a Calabi--Yau has $SU(3)$-structure, and it is known that there are no $SU(3)$-structure solutions with localised $O$-planes \cite{Acharya:2006ne}. Therefore, any solution must deform the metric away from the Ricci flat one. For this deformation to be supersymmetric it should exhibit $SU(3)\times SU(3)$ structure, which is the most general possibility. We therefore study whether there are ten-dimensional solutions with $SU(3)\times SU(3)$-structure that exhibit the properties of DGKT. Even though these would not be compactifications on the Ricci-flat Calabi--Yau metric, they may morally be considered the uplifts to DGKT. 

We find an approximate solution to the supersymmetry equations. Specifically, in \cite{Saracco:2012wc} it was proposed that one could look for solutions that are perturbations of the smeared solution controlled by an expansion parameter related to the value of the cosmological constant. Following this approach, we find a solution to the supersymmetry equations and the Bianchi identities with localised sources, at leading order in this expansion parameter. This is the main result of the paper. The solution is very different to the one considered in \cite{Saracco:2012wc}, specifically we have an exactly, rather than approximately, vanishing Freund--Rubin flux.\footnote{It should be noted that AdS solutions without a Freund--Rubin flux exist, notably in cases where they are forbidden by dimensionality of spacetime, such as for AdS$_5$ in M-theory \cite{lin-lunin-maldacena} or AdS$_7$ in type IIA \cite{afrt}, but also not, such as in \cite{Couzens:2016iot}.}  However, the methodology is the same.\footnote{An argument against exactly vanishing Freund--Rubin flux was suggested in \cite[Sec.~7.5]{Saracco:2012wc}. However, we have found a mistake in that argument (which does not influence the rest of the paper).} 

Note that this approach was also recently utilised in a closely related paper \cite{Junghans:2020acz} (which appeared as this paper was nearing completion.) Our work focuses on the supersymmetry equations, which were not considered in \cite{Junghans:2020acz}, but at least so far as the existence of a first-order solution our results agree with those of \cite{Junghans:2020acz}. 

Returning to the question of separation of scales, our results show that DGKT has passed a first non-trivial test. However, we do not claim that our results show conclusively that DGKT really does uplift to a full exact solution of string theory, nor that if such a solution exists it exhibits separation of scales. We discuss the remaining open questions 
in section \ref{sec:conclu}.

The paper is organised as follows. In section \ref{sec:IIA} we review the basics ingredients of DGKT that are important for a 10d description, and the most general class of 10d supersymmetric backgrounds that they can correspond to. In section \ref{sec:nogosol} we discuss how the 4d features of DGKT constrain such 10d supersymmetric vacua, narrowing down the search for solutions. In section \ref{sec:approx} we present a large volume/weak coupling approximation of the supersymmetry equations compatible with DGKT. In section \ref{sec:Bianchi} we solve exactly the Bianchi identities  corresponding to DGKT in a generic Calabi--Yau. In section \ref{sec:solution} we present our solution to the supersymmetry equations and Bianchi identities in the large volume approximation. We express such a solution in terms of Calabi--Yau quantities, and discuss its general features. We finally draw our conclusions in section \ref{sec:conclu}.

The most technical details of the paper have been relegated to the Appendices. Appendix \ref{ap:SU33} reviews type IIA supersymmetry equations from the viewpoint of $SU(3)\times SU(3)$ structures, and appendix \ref{ap:SBEproof} contains the proof of what we dub the source balanced equation, see \eqref{preGOE}.

\section{Supersymmetric type IIA flux vacua}
\label{sec:IIA}

In this section we review the setup considered in \cite{DeWolfe:2005uu}, and in particular the features that should appear in a 10d description. Since the vacua found in \cite{DeWolfe:2005uu} are supersymmetric from a 4d viewpoint, one expects their corresponding 10d backgrounds to solve the 10d supersymmetry equations with four supercharges. These equations can be efficiently encoded in the language of compactifications with $SU(3)\times SU(3)$ structures \cite{Grana:2005sn}, which we also review. As we will show in the next section, the results of \cite{DeWolfe:2005uu} imply that only a specific class of $SU(3)\times SU(3)$-structure compactifications can describe the global aspects of these vacua. 

\subsection{4d description of type IIA AdS$_4$ orientifold vacua}

Let us consider type IIA string theory compactified in an orientifold of $X_4 \times X_6$ with $X_6$ a compact real six-manifold with a Calabi--Yau metric, and therefore a K\"ahler 2-form $J_{\rm CY}$ and a holomorphic 3-form $\Omega_{\rm CY}$. Following the standard construction \cite{Ibanez:2012zz}, we take the orientifold action to be generated by $\Omega_p (-1)^{F_L}\cR$,\footnote{Here $\Omega_p$ is the worldsheet parity reversal operator, ${F_L}$ is the space-time fermion number for the left-movers.} with $\cR$ an anti-holomorphic involution acting as $\cR J_{\rm CY}=-J_{\rm CY}$ and $\cR\Omega_{\rm CY} = - \overline{\Omega}_{\rm CY}$. The fixed locus $\Pi_{\rm O6}$ of $\cR$ is one or several 3-cycles of $X_6$ in which O6-planes are located. In a consistent compactification, the RR charge of such O6-planes must be cancelled by a combination of D6-branes wrapping three-cycles of $X_6$ and background fluxes. 

To describe the set of background fluxes one may use the democratic formulation of type IIA supergravity \cite{Bergshoeff:2001pv},  in which all RR potentials are grouped in a polyform ${\bf C} = C_1 + C_3 + C_5 + C_7 + C_9$, and so are their gauge invariant field strengths
\be
{\bf G} \,=\, d_H{\bf C} + e^{B} \wedge {\bf \bar{G}} \, ,
\label{bfG}
\ee
with $H$ the three-form NS flux, $d_H \equiv (d - H \wedge)$ is the $H$-twisted differential  and ${\bf \bar{G}}$ a formal sum of closed $p$-forms of $X_6$. The Bianchi identities read
\begin{equation}\label{IIABI}
\ell_s^{2} \,  d (e^{-B} \wedge {\bf G} ) = - \sum_\a \lambda \left[\delta (\Pi_\alpha)\right] \wedge e^{F_\alpha} \, ,  \qquad d H = 0 \, ,
\end{equation} 
with $\ell_s  =  2\pi \sqrt{\a'}$ the string length. Here $\Pi_\alpha$ hosts a D-brane source with a quantised worldvolume flux $F_\alpha$, and $\delta(\Pi_\alpha)$ is the bump $\delta$-function form with support on $\Pi_\alpha$ and indices transverse to it, such that $\ell_s^{p-9} \d(\Pi_\a)$ lies in the Poincar\'e dual class to $[\Pi_\a]$. O-planes contribute as D-branes but with minus four times their charge and $F_\alpha \equiv 0$.  Finally $\lambda$ is the operator that reverses the order of the indices of a $p$-form. Page charge quantisation reads \cite{Marolf:2000cb}
\begin{equation}
\frac{1}{\ell_s^{p}} \int_{\pi_{p+1}} \left[ d(e^{-B} \wedge {\bf C}) + \bar{\bf G} \right]_{p+1}  \in  \IZ\, , \qquad  \frac{1}{\ell_s^{2}}\int_{\pi_3}  H \in \IZ\, ,
\label{Pqu}
\end{equation} 
where $\pi_{p+1} \in X_6$ stands for any $(p+1)$-cycle not intersecting the $\Pi_a$'s. In the absence of localised sources, $e^{-B} \wedge {\bf C}$ is globally well-defined and the $\bar{G}_{p+1}$ are quantised, so one can define the internal RR flux quanta in terms of the following integer numbers
\begin{equation}
m \, = \, \ell_s \bar{G}_0\, ,  \quad  m^a\, =\, \frac{1}{\ell_s^5} \int_{X_6} \bar{G}_2 \wedge \tilde \om^a\, , \quad  e_a\, =\, - \frac{1}{\ell_s^5} \int_{X_6} \bar{G}_4 \wedge \om_a \, , \quad e_0 \, =\, - \frac{1}{\ell_s^5} \int_{X_6} \bar{G}_6 \, ,
\label{RRfluxes}
\end{equation}
with $\om_a$, $\tilde \om^a$ integral harmonic two- and four-forms such that $\ell_s^{-6} \int_{X_6} \om_a \wedge \tilde{\om}^b = \delta_a^b$. From the 4d EFT viewpoint, \eqref{RRfluxes} enter the flux generated superpotential for the would-be K\"ahler and B-field moduli of the compactification \cite{Taylor:1999ii}, dominating their dynamics in the large volume regime. 

In the presence of only O6-planes the Bianchi identities for the RR fluxes read
\be
dG_0 = 0\, , \qquad d G_2 = G_0 H + \d_{\rm O6}\, ,  \qquad d G_4 = G_2 \wedge H\, , \qquad dG_6 = 0\, ,
\label{BIG}
\ee
where  we have defined $\d_{\rm O6}\equiv - \ell_s^{-2}  4 \d(\Pi_{\rm O6})$. This in particular implies that
\be
4 {\rm P.D.} [\Pi_{\rm O6}] = m [ \ell_s^{-2} H] \, ,
\label{tadpole}
\ee
constraining the quanta of Romans parameter and NS flux. 

Ref. \cite{Grimm:2004ua} obtains a 4d effective F-term potential by combining {\it i)} the classical K\"ahler potential of Calabi--Yau orientifolds without fluxes and {\it ii)} the superpotential generated by the RR flux quanta \eqref{RRfluxes} and NS three-form flux quanta. The approach in \cite{DeWolfe:2005uu} performs a 4d analysis of such potential,  finding an infinite discretum of $\cN=1$ AdS$_4$ vacua. Interestingly, this features of such vacua can be easily expressed in terms of integrals of 10d gauge invariant field strengths, which in 4d language are seen as specific combinations of flux quanta and axionic scalars \cite{Bielleman:2015ina,Carta:2016ynn,Herraez:2018vae}. Indeed, one finds that (see e.g. \cite{Marchesano:2019hfb})
\be
\langle g_s^{-1}\rangle [ H ]  = \frac{2}{5} G_0 [\re \Om_{\rm CY} ] \, , \qquad \langle G_2\rangle  =  0\, ,  \qquad \langle G_4\rangle  =   \frac{3}{10}G_0 \, , \qquad \langle G_6\rangle  =  0\, , 
\label{intflux}
\ee
where we have defined
\be
\langle g_s^{-1} \rangle = \frac{\int_{X_6} e^{-\phi}  J_{\rm CY}^3 }{\int_{X_6} J_{\rm CY}^3}\, , \ 
\langle G_2 \rangle = \frac{\int_{X_6} G_2 \wedge J_{\rm CY}^2 }{\int_{X_6} J_{\rm CY}^3}\, , \ \langle G_4 \rangle = \frac{\int_{X_6} G_4  \wedge J_{\rm CY}}{\int_{X_6} J_{\rm CY}^3 }\, , \ \langle G_6 \rangle = \frac{\int_{X_6} G_6}{-\int_{X_6} J_{\rm CY}^3}\, .
\ee
with $\phi$ the 10d dilaton.\footnote{Remarkably, similar relations hold when adding curvature corrections \cite{Palti:2008mg,Escobar:2018rna} and mobile D6-branes \cite{Escobar:2018tiu}.} It is also obtained that the Calabi--Yau volume $\ell_s^6 \cV_{\rm CY} (X_6) = -\frac{1}{6} {\int_{X_6} J_{\rm CY}^3}$ is controlled by the four-form and two-form flux quanta, more precisely by the combination $\hat{e}_a = e_a - \oh \frac{\cK_{abc} m^am^b}{m}$, with $\cK_{abc}$ the triple intersection numbers of $X_6$ \cite{DeWolfe:2005uu}. As such, one may arbitrarily increase the Calabi--Yau volume by increasing the value of $\hat{e}_a$, while the density of four-form flux remains constant, as captured by \eqref{intflux}. On the other hand, because of \eqref{tadpole} the quanta of $H$ and $m$ are bounded and in practice be considered to be fixed. This implies that the average value of the (inverse) dilaton scales as
\be
\langle g_s^{-1}\rangle \, \sim \, \cV_{\rm CY}^{1/2} \, \sim\, \hat{e}^{3/4}\, .
\label{scaling0}
\ee
Finally, the AdS$_4$ radius $R_{\rm AdS}$ scales like 
\be
R_{\rm AdS} M_{\rm P} \sim \hat{e}^{9/4}\, .
\label{4dRads}
\ee
Therefore the 4d EFT considered in \cite{DeWolfe:2005uu} suggests that as we increase $\hat{e}$ along the infinite family of solutions we go to a limit of weak coupling, large volume and large AdS radius.

\subsection{10d description and supersymmetry equations} 
\label{sub:10d-susy}

We now review the conditions for ten-dimensional supersymmetry, in the pure spinor formalism. 
A ten-dimensional AdS$_4$ vacuum has a metric of the form
\begin{equation}\label{eq:warped-product}
	ds^2 = e^{2A}ds^2_{\mathrm{AdS}_4} + ds^2_6\,.
\end{equation}
As in the previous subsection, we consider only internal fluxes $G_k$; the external fluxes are determined by duality. 

Preserved supersymmetry  imposes differential equations on the internal part of the supersymmetry parameters $\eta^a_\pm$. From these one can build a bispinor $\Phi_\pm \equiv \eta^1_+ \otimes  \eta^{2\,\dagger}_\pm$, which can be interpreted as a polyform in the internal space by the Clifford map $\gamma^m \to d x^m$. This form obeys some algebraic constraints, that follow from its definition in terms of spinors, and some differential equations that follow from supersymmetry. 

The algebraic conditions allow several types of solutions. Only two classes are relevant for us. The first class is made of the $SU(3)$-structure solutions, and are enough to describe the smeared uplift of DGKT in \cite{Acharya:2006ne}; they depend on a two-form $J$ and a three-form $\Omega$. The second class, which is the generic solution, comprises the $SU(3)\times SU(3)$-structure solutions; this is the one relevant for this paper. Both classes are reviewed in App.~\ref{ap:SU33}; here we only need to know that the $SU(3)\times SU(3)$ class depends on the following data:
\begin{itemize}
	\item Two functions $\psi$, $\theta$, 
	\item A complex one-form $v$,
	\item A real two-form $j$,
	\item A complex two-form $\omega$. 
\end{itemize}
The function $\psi$ measures the departure from the $SU(3)$ class; $\psi\to 0$ makes one fall to that case. In that limit, the data reassemble in those of an $SU(3)$-structure as
\begin{equation}\label{eq:JO-jo}
	J = j + \frac i{2\tan^2\psi} v \wedge \bar v \, ,\qquad \Omega = \frac i{\tan\psi}\, v \wedge \omega \,.
\end{equation}
On the other hand, $\theta$ is the phase of $\eta^{2\,\dagger}_+\eta^1_+$.   Finally, the forms $v$, $j$, $\omega$ are the same data that define an $SU(2)$-structure in six dimensions, see App.~\ref{ap:SU33} for more details.

The differential equations imposed by supersymmetry on the $\Phi_\pm$ can be written as \cite{Grana:2005sn}
\begin{subequations}\label{eq:psp}
	\begin{align}
		\label{eq:psp+}
		d_H \Phi_+ &= - 2 \mu e^{-A} \mathrm{Re} \Phi_-\, , \\
		\label{eq:psp-}
		d_H \left(e^{A} \im{ \Phi_-}\right) &= -3 \mu \mathrm{Im} \Phi_+ + e^{4A}\star \lambda \mathbf{G}\,.
	\end{align}
\end{subequations}
Here $\star$ is the internal Hodge dual, and $\lambda$ is a sign reversal operation defined on a $k$-form as $\lambda(\alpha_k)= (-1)^{\lfloor k/2 \rfloor} \alpha_k$. The mean value of $e^{-A}$ can fixed by shifting $A$ by a constant and absorbing its effect into the definition of $\mu\equiv \sqrt{-\Lambda/3}$, which is the AdS$_4$ scale seen from the 10d string frame perspective (not to be confused with $1/R_{AdS}$ in \eqref{4dRads}, even if related to it).
For our purposes, it is more convenient to replace the second by the alternative expression \cite{Tomasiello:2007zq}
\begin{equation}\label{eq:calJ}
	{\mathcal J}_+ \cdot d_H \left( e^{-3A} \im{\Phi_-}\right) = -5 \mu e^{-4A} \mathrm{Re} \Phi_+ + \mathbf{G}\,.
\end{equation}
The new operator ${\mathcal J}_+\cdot$ is associated in a certain way to the form $\Phi_+$; we will see in explicit examples what it reduces to. 

Let us now present in some detail what one gets by plugging in (\ref{eq:psp+}), (\ref{eq:calJ}) the solutions to the algebraic constraints for $\Phi_\pm$. Here we will focus on those classes of solutions that are more relevant for the computations of the following sections, leaving the rest for the more detailed discussion of App.~\ref{ap:SU33}.

\subsubsection{$SU(3)$-structure} 
\label{ssub:su3}

For an $SU(3)$-structure the pure spinors have the form
\be \label{eq:Phi-su3}
\Phi_+ \, =\, e^{3A-\phi} e^{i\theta} e^{-i J}\, , \qquad \qquad \Phi_-\,  =\, e^{3A-\phi}\Omega \, .
\ee
where $J$ and $\Omega$ do not need to be closed, allowing for $SU(3)$-structure torsion classes. From here one finds that 
\begin{equation}
 d\th =0\,\qquad 3dA =d\phi   
\end{equation}
and the following expression for the fluxes \cite{Koerber:2010bx}:
\begin{subequations}	
\label{su3flux}
\begin{align}
H & = 2 \mu e^{-A} {\rm cos}\, \th\, \re \Om\, ,\\
\label{su3fluxG0}
G_0 & = 5 \mu e^{-\phi-A}  {\rm cos}\, \th\, , \\
\label{su3fluxG2}
G_2 & = \frac{1}{3} \mu e^{-\phi-A}  {\rm sin}\, \th\, J - J \cdot d \left( e^{-\phi}  \im  \Om \right)\, , \\
G_4 & =  \frac{3}{2} \mu e^{-\phi-A}  {\rm cos}\, \th\, J \wedge J \, , \\
G_6 & = 3 \mu e^{-\phi-A}  {\rm sin}\, \th\, d\mathrm{vol}_{X_6}\, .
\end{align}
\end{subequations}     
The operation $J\cdot$ is defined as $J^{-1} \llcorner:$ one inverts the two-form $J$ to obtain a bivector, and one contracts this bivector with the forms that follow it. We will see more precisely how that works in the solutions below.

\subsubsection{$SU(3)\times SU(3)$ with $\theta=0$} 
\label{ssub:t0}

Here we consider the special case $\theta=0$, since, as we will argue in section \ref{sec:nogosol}, this case is the one suitable for a microscopic description of the DGKT vacua. $SU(3)\times SU(3)$ backgrounds with $\theta\neq 0$ have a similar but slightly more involved description; we defer their discussion to App.~\ref{sub:tneq0}.

In the case $\theta=0$ the pure spinors have the form
\begin{equation}
	\label{su3phi}
		\Phi_+ = e^{3A-\phi} \cos\psi \exp[-i J_\psi] \, ,\qquad \Phi_- =e^{3A-\phi} \cos\psi\, v \wedge \exp[i \omega_\psi]\,,
	\end{equation}
where
\begin{equation}\label{eq:Jpsi}
	J_\psi \equiv \frac1{\cos(\psi)}j + \frac i{2 \tan^2(\psi)} v\wedge \bar v \,,\qquad 
	\omega_\psi \equiv \frac1{\sin(\psi)} \left({\rm Re} \omega + \frac i{\cos(\psi)} {\rm Im} \omega\right)\,.
\end{equation}    
Details about $v$, $j$, $\omega$ are reviewed in App.~\ref{ap:SU33}.

There are first some equations that do not involve the fluxes:
\begin{subequations}	
\begin{align}
	\label{eq:revt0}    
	&{\rm Re} v= - \frac{e^A}{2 \mu}\left(3 dA - d \phi - \tan \psi d \psi\right) = -\frac{e^A}{2\mu} d\log(\cos\psi e^{3A-\phi}) \, ,\\
    \label{eq:dJpsit0}
    &d(e^{3A-\phi} \cos \psi J_\psi)=0 \, .
\end{align}
\end{subequations}

To arrive to \eqref{su3phi} one needs to perform a B-field transformation on the pure spinors and the fluxes \cite{gualtieri,Grana:2006kf,Saracco:2012wc}, with $b_{\Phi_\pm} = {\rm tan}\, \psi\, \im \om$. The physical fluxes are obtained by undoing it:
\begin{equation}\label{eq:GF-t0}
	H= \hat H + d(\tan\psi \mathrm{Im} \omega) \, ,\qquad \mathbf{G}= e^{\tan \psi \mathrm{Im} \omega \wedge} \mathbf{F}\,,
\end{equation}
where
\begin{subequations}\label{eq:flux-t0}
\begin{align}
    \label{eq:Ht0}
 	& \hat H= 2\mu e^{-A} {\rm Re} (i v \wedge \omega_\psi) \, , \\
	\label{eq:F0t0}
	& F_0 = -J_\psi\cdot
    d (\cos \psi e^{-\phi}{\rm Im} v)
    + 5 \mu \cos \psi e^{-A-\phi} \,,\\  
	\label{eq:F2t0}
	& F_2 = -J_\psi\cdot d\, {\rm Im} (i \cos \psi e^{-\phi} v \wedge \omega_\psi) - 2 \mu \frac{\sin^2 \psi}{\cos \psi} e^{-A-\phi} {\rm Im} \omega_\psi\,,\\
	\label{eq:F4t0}
& F_4 = J_\psi^2\left[ \frac12 F_0 - \mu \cos \psi e^{-A-\phi}\right] + J_\psi \wedge d\, {\rm Im} (\cos \psi e^{-\phi} v) \,,\\
& F_6 = 0 \label{f6su3su3}\,.
\end{align}
\end{subequations}
In the limit $\psi \rightarrow 0$ and upon making the replacements \eqref{eq:JO-jo} one recovers \eqref{su3flux} with $\theta=0$.

\section{Constraining the solution}
\label{sec:nogosol}

As expected for data obtained from the 4d EFT, the relations \eqref{intflux} correspond to integrated quantities, and so there could be an infinite number of 10d backgrounds that correspond to them. Nevertheless, when combined with supersymmetry they result in some stringent constraints on the microscopic description of DGKT vacua. In this section we develop such  constraints by using the machinery of $SU(3)\times SU(3)$-structure compactifications. The result is quite simple to state: DGKT vacua should correspond to 10d backgrounds such that {\it i)} the internal flux $G_6$ vanishes pointwise and {\it ii)} it corresponds to a genuine $SU(3)\times SU(3)$ structure with $\theta=0$. 

\subsection{The Freund--Rubin flux}
\label{sec:susFR}

A key characteristic of type IIA flux compactifications studied so far is their Freund--Rubin flux. Given the compactification Ansatz \eqref{eq:warped-product}, this flux is of the form
\be
G^{\rm{10d}}_4 =  c\,  d\mathrm{vol}_{X_4} + G_4 \, ,
\ee
where $c$ is defined by
\be
c = e^{4A} \star G_6\, ,
\ee
and $G_4$, $G_6$ are the four- and six-form components of the internal flux \eqref{bfG}. The equations of motion imply that $c$ is a constant, since
\be
d\left(\star_{10} G_6\right) =  d\mathrm{vol}_{X_4} \wedge d \left(e^{4A} \star G_6 \right) =  d\mathrm{vol}_{X_4} \wedge d c = 0 \, ,
\ee
or in other words that the internal six-form takes the expression
\be
G_6 = c\, e^{-4A} d\mathrm{vol}_{X_6} \, ,
\ee
with $c$ constant.
Notice that the volume form $d\mathrm{vol}_{X_6}$ need not be $-\frac{1}{3} J_{\rm CY}^3$, because the actual internal metric of the solution is not supposed to be Calabi--Yau, even if $X_6$ admits a Calabi--Yau metric. In any case the last relation in \eqref{intflux} reads
\be
c \int_{X_6} e^{-4A} d\mathrm{vol}_{X_6} = 0 \, .
\label{inte4a}
\ee
This in principle leads to two possibilities: either $c$ or the integral vanishes. Notice however that the integrand is positive definite -- excluding perhaps regions very close to the O6-planes where the supergravity approximation breaks down -- and so should be its integral.\footnote{In practice one may shifts the warp factor by a constant that is absorbed into the AdS$_4$ scale $\mu$, to fix the value of the integral to a certain positive value. We take the simple choice $\langle e^{-4A}\rangle = 1$ in our solution of section \ref{sec:solution}.} Therefore, sensible 10d uplifts of DGKT vacua are those in which the six-form flux $G_6$ (and dual Freund--Rubin flux) must vanish point-wise on $X_6$ 
\be
G_6 = 0 \, .
\label{f6ha0}
\ee
As follows from the discussion of Appendix \ref{ap:SU33}, this condition has a straightforward implication for the poly-forms describing the $SU(3)\times SU(3)$ structure. Namely
\be
\left. \im \Phi_+ \right|_{\mathrm{0-form}} =0\, .
\label{imphi0}
\ee
This simple constraint rules out several candidates for DGKT 10d vacua.

\subsection{No $SU(3)$-structure solution}
\label{sec:nosu3}

As noticed in \cite{Acharya:2006ne}, the relations \eqref{intflux} are very suggestive from the viewpoint of type IIA flux backgrounds with $SU(3)$ structure, whose most general solution was found in \cite{Behrndt:2004km,Lust:2004ig}. Nevertheless, this particular subcase of $SU(3)\times SU(3)$-structure compactification cannot accommodate a 10d uplift of \cite{DeWolfe:2005uu} unless the orientifold sources are smeared. While this is a well-known obstruction, it will prove useful to review it from the present perspective. 

Recall the $SU(3)$-structure solutions (\ref{su3flux}). It is easy to see that the choice $\theta=0$ is reminiscent of the relations \eqref{intflux}, and in particular that it is compatible with the constraints \eqref{f6ha0} and \eqref{imphi0}. However, this choice is not allowed in the present setup, unless the Bianchi identity is modified by smearing the O6-plane sources. Indeed, it follows from the Bianchi identity for $G_0$ that there both the warp factor and dilaton are constant, from where one obtains that 
\be
d  \im  \Om = i  W_2 \wedge J\, ,
\ee
with $W_2$ a real, primitive (1,1)-form. Then, the Bianchi identity for $G_2$ becomes \cite{Acharya:2006ne,Koerber:2007jb}
\be
e^{-\phi} \left[ \frac{1}{4} |W_2|^2 + e^{-2A} \mu^2 \left( 10\,  {\rm cos}^2 \th - \frac{2}{3}  {\rm sin}^2 \th\right)  \right]  \re \Om = - \delta_{O6}\, .
\label{tf2loc}
\ee
Away from the O6-plane locus the lhs of \eqref{tf2loc} needs to vanish, which necessarily imposes that $\theta \neq 0$ and a non-vanishing internal flux $G_6$. Therefore, by the requirement \eqref{f6ha0} this cannot be a 10d realisation of \cite{DeWolfe:2005uu}. If $\d_{O6}$ is replaced with a smeared three-form source in the appropriate cohomology class
\be 
\label{eq:smearing}
- \delta_{O6}\, \rightarrow \, G_0H = 10 e^{-\phi-2A} \mu^2   {\rm cos}^2 \th\,  \re \Om 
\ee
such obstruction is gone, and one find solutions with $W_2 = \theta =0$. This would-be solution would have $dJ = d\Omega =0$, and would correspond to a Calabi--Yau metric. The fluxes would read
\begin{equation}\label{eq:SU3th0}
	\begin{split}
		H&= 2 \mu {\rm Re}\, \Omega\, , \\
		G_0&= 5 \mu e^{- \phi} \, ,\qquad  \quad \ G_2 =0\ ,\\
		G_4&= \frac32 \mu e^{- \phi} J^2 \, ,\qquad G_6 = 0 \, . 
	\end{split}	
\end{equation}
Since $A$ is constant, we have set it to zero, reabsorbing it in $\mu$. To see how things scale, assume as in \cite{DeWolfe:2005uu} that $F_0\sim O(1)$ and that the internal space has volume $\cV(X_6) \sim R^6$ in string units. We know already that $\delta \propto {\rm Re}\,  \Omega$; if we take $\delta_{O6} \sim - \frac 1 {R^3} {\rm Re}\, \Omega$, integrating $\delta$ along a 3-cycle gives $O(1)$, as it should. From all this we read
\begin{equation}\label{eq:scales}
	g_s = \frac{5}{m} \hat{\mu} \sim R^{-3}\,,
\end{equation}
with $\hat{\mu} = \mu\ell_s$, in agreement with \eqref{scaling0}.

It has been recently proposed in \cite{Junghans:2020acz} that this Calabi--Yau solution with smeared sources can be seen as the leading order contribution to an expansion in the flux quantity $\hat{e}$ controlling the volume of the compactification. As we will see in section \ref{sec:solution}, this is manifest for the solution that we find for the $SU(3)\times SU(3)$-structure supersymmetry equations, approximated in the large volume regime. Before deriving such equations, let us constrain which kind of $SU(3)\times SU(3)$ structure can describe DGKT microscopically.

\subsection{Setting $\theta = 0$}
\label{sec:nothetaneq0}

Massive type IIA supergravity backgrounds solving the $SU(3)\times SU(3)$-structure supersymmetry equations have been analysed in \cite{Gaiotto:2009mv,Saracco:2012wc}. In particular, in \cite{Saracco:2012wc} two different branches of solutions were identified, as reviewed in section \ref{sub:10d-susy} and Appendix \ref{ap:SU33}. These two branches are described in terms of the function $\theta$ defined in section \ref{sub:10d-susy}, which  appears in the pure spinors as in \eqref{ap:su3phi}. One branch has $\theta=0$ (see section \ref{sub:10d-susy}) and the other one has non-vanishing, varying $\theta$ (see section \ref{sub:tneq0}).

Our discussion above implies that the branch with $\theta=0$ should be more suitable to describe DGKT. Indeed, given \eqref{ap:su3phi} one can rewrite \eqref{imphi0} as $\theta = 0$ or $\pi$. Accordingly, one can show that after the B-field transformation \eqref{eq:GF-t0} one obtains $G_6=0$ from \eqref{eq:flux-t0}, see Appendix \ref{ap:SU33}.  The compatibility of this branch with the presence of O6-planes seemed unlikely from the symmetry arguments used in \cite{Saracco:2012wc}. However, in the following sections will see that supersymmetry equations for the case $\theta=0$ are rich enough to host localised and smeared sources at the same time. In this sense what our results of the following sections suggest is that the 10d description of \cite{DeWolfe:2005uu} consists of a $SU(3)\times SU(3)$-structure background with $\theta=0$ that at large volumes can be approximated by an $SU(3)$-structure background with $\theta=0$. Indeed, we will see that the background that we find can be organised as a perturbative expansion on the small parameter $g_s \sim \cV_{X_6}^{-1/2}$. The zeroth order contribution is nothing but the background \eqref{eq:SU3th0}.

The branch in which $\theta$ is non-vanishing, is a priori not suitable  to describe 10d uplift of DGKT. First, as reviewed in section \ref{sub:tneq0}, such a solution has a varying $G_6$ flux, which prevents it to satisfy the point-wise constraint \eqref{f6ha0}. In addition, this sort of backgrounds are characterised by an NS three-form flux $H$ which is exact. As this implies vanishing $H$-flux quanta, it can never describe the global features of a DGKT vacuum. Finally, a crucial aspect of this branch is that $\left. \im \Phi_+ \right|_{\mathrm{0-form}} \neq 0$, and in fact it is not even constant. The aim of \cite{Saracco:2012wc} was to find a solution which only asymptotes to $\left. \im \Phi_+ \right|_{\mathrm{0-form}} = 0$, but we have shown that this must hold locally anywhere on $X_6$ where a 10d supergravity description is reliable. Therefore it seems unlikely the solution in \cite{Saracco:2012wc} can be part of a 10d description of  DGKT.

This being said, let us stress that the approximate solution that we find in section \ref{sec:solution} breaks down near the O6-plane loci. In particular in those regions we find the same metric singularities associated with O6-planes in flat space, featuring a divergent negative warp factor $e^{-4A}$. In the case of flat space it is known that the divergent negative warp factor around the O6-plane is resolved by string theory corrections, uplifting the solution to M-theory on a $G_2$ manifold with an Atiyah-Hitchin metric on the former O6-plane region \cite{Gibbons:1986df,Sen:1997kz,Seiberg:1996nz,Hanany:2000fw}. In the case at hand we are dealing with massive type IIA string theory and therefore we lack an M-theory description, so it would be very interesting to understand how the theory resolves such a singularity. One possibility could be that the full solution with $\theta =0$ does not have any singularity. This would be quite analogous to the result found in \cite{Saracco:2012wc} for the $\theta \neq0$ branch. Indeed, there it was shown that such massive type IIA solutions with O6-planes can resolve the O6-plane singularity  without resorting to an M-theory description. As these belong to a different branch of $SU(3)\times SU(3)$-structure backgrounds, we will take an agnostic approach towards this possibility.

\subsection{The source balanced equation}
\label{sec:SBE}

If the obstruction for $SU(3)$-structure solutions can be circumvented by $SU(3)\times SU(3)$-structure backgrounds with $\theta=0$ a natural question is how the equation \eqref{tf2loc} leading to the obstruction is modified. In the following we would like to present a generalisation of eq.\eqref{tf2loc}, valid for any $SU(3)\times SU(3)$-structure background, which we dub {\em source balanced equation}. 

For this we first need to introduce the Mukai pairing
\be
\left< \omega_1 , \omega_2 \right> \equiv \left. \omega_1 \wedge \lambda \left( \omega_2 \right) \right|_{\mathrm{top}}\, ,
\ee
for the poly-forms $\omega_1$ and $\omega_2$, where $|_{\rm top}$ indicates that we only extract the top form of the product. The source balanced equation then reads
\be
3\mu^2 e^{-4A} \left< \mathrm{Re\;}\Phi_+ , \mathrm{Im\;}\Phi_+ \right>  - e^{4A} \sum_k G_k \wedge \star G_k + dX_5 
=  \left< \delta^{(3)}_{\mathrm{source}} , e^A  \mathrm{Im\;}\Phi_- \right> \; ,
 \label{preGOE}
\ee
where $\delta^{(3)}_{\mathrm{source}} = \sum_\a \delta (\Pi_\alpha)$ contains all the localised sources wrapping three-cycles, both O6-planes and D6-branes. In addition
\be
X_5 \equiv \left<  e^A \mathrm{Im\;}\Phi_-, {\bf G} \right> =   -G_2 \wedge \left( e^A \mathrm{Im\;}\Phi_- \right)_{3}  + G_4 \wedge \left( e^A \mathrm{Im\;}\Phi_- \right)_{1} + G_0 \left( e^A \mathrm{Im\;}\Phi_- \right)_{5}   \;,
\label{X5}
\ee
where the subscripts denote the degree of the form to be picked out. 

Eq.\eqref{preGOE} is derived in Appendix \ref{ap:SBEproof} by using the Bianchi identities and the supersymmetry equations. Notice that it takes a similar form to \eqref{tf2loc} in the sense that the left-hand side is supported over the whole manifold, while the right-hand side is localised. One can see this relation as a generalisation of (\ref{tf2loc}), in which case one had $X_5=0$. Indeed, recall that in the SU(3)-structure case $ \left( \mathrm{Im\;}\Phi_- \right)_{1} = \left( \mathrm{Im\;}\Phi_- \right)_{5} = 0$ and that $G_2$ is a (1,1)-form. In section \ref{ss:comparison} we will test our solution  with this equation, to see in which way \eqref{tf2loc} is modified to allow for a consistent $SU(3)\times SU(3)$-structure solution.

\section{The large volume approximation}
\label{sec:approx}

We now consider the BPS equations in a limit where the cosmological constant is small, aiming for a solution similar to (\ref{eq:SU3th0}) but \emph{without smearing}. We will do this by taking the parameter $\hat\mu = \mu\ell_s$ in (\ref{eq:psp}), (\ref{eq:calJ}) to be small; recalling (\ref{eq:scales}) $g_s$ will then also be small and $R$ large. 

\subsection{Defining the limit} 
\label{sub:def-lim}

As we have seen in section \ref{sec:nogosol}, the smeared solution comes from an $SU(3)$-structure, but the solution with localised O6-planes that we are looking for cannot. As discussed around (\ref{eq:JO-jo}), the function $\psi$ interpolates between $SU(3)$ and $SU(3)\times SU(3)$. So in the limit we also have to take the function $\psi$ to be at least of order $\hat\mu \sim R^{-3}\sim g_s$ at leading order, recalling (\ref{eq:scales}). In addition, from (\ref{eq:SU3th0}) we see that for $F_0$ to stay non-zero in the limit we need $e^\phi \to 0$. On the other hand, since we are already making $\hat\mu\to0$, $e^A$ should not scale. For simplicity in the following we will  fix $\langle e^{-4A}\rangle = 1$.

A limit with all these features was originally devised in \cite{Saracco:2012wc}, exactly for the solution at hand. As we commented earlier, there the focus was on the local behaviour, and we have argued above that solution cannot capture the global solution, essentially because $\theta\neq 0$ was taken there. In the limit, the problem presented itself already in eq.\cite{Saracco:2012wc}; it was noted below eq.(6.12) in that paper that $\mathrm{Re}\,  \Omega$ has to be exact at leading order, and that this could be an obstruction to finding a global solution.

Nevertheless, we can still apply the same ideas of \cite{Saracco:2012wc} to the $\theta=0$ case. Notably, it was decided there to expand in $\mu$, but with a subleading behaviour that is either an even or odd function of $\mu$. This was found to simplify the equations significantly, and it was inspired in turn by a similar limit in \cite{Gaiotto:2009yz} where $\psi \to 0$ but the cosmological constant remained fixed. 

Implementing this strategy in our case leads us to taking $g_s\to 0$, with the following Ansatz: 
\begin{equation}\label{eq:limit}
\begin{split}
	&\hat{\mu} = \frac m5 g_s\, ,\qquad \psi = g_s \psi_1 +  \cO(g_s^3) \, ,\qquad \theta=0 \, ,\\
	&e^\phi = g_s e^{\delta\phi_0 + g_s^2 \delta\phi_2 + \cO(g_s^4)}   \, ,\qquad e^A = e^{A_0 + g_s^2 A_2 + \cO(g_s^4)}\,.
\end{split}	 
\end{equation}

It is important to stress that the equations will fix the coefficients of the expansion as a function of the coordinates, in such a way that some extra powers of the parameter $g_s\sim R^{-3}$ will appear. For example we will find below that
\begin{equation}
    e^{A_0}\sim a_0 + a_1 R^{-4} \sim a_0 + a_1 g_s^{4/3}\,.
\end{equation}
This might look confusing, but the method is sensible as long as these ``hidden'' powers of $g_s$ are not smaller with respect to terms we have ignored in (\ref{eq:limit}). The same comment applies to the expansion of the forms, to which we now turn.

\subsection{Forms and fluxes} 
\label{sub:lim-forms}

We now have to decide how to scale the forms. $\mathrm{Re}\, v$ is already determined by (\ref{eq:revt0}). Due to the $\mu$ in the denominator of that equation,  (\ref{eq:limit}) would imply that $\mathrm{Re} v \sim -\frac{5e^A}{2g_s m}d(3A_0- \delta \phi_0)$, whereas as we explained above we would like $\mathrm{Re}\, v\to 0$ in the limit. For this reason we take 
\begin{equation}
    \delta \phi_0 = 3 A_0\,.
\end{equation}
Now we obtain
\begin{equation}\label{eq:Rev1-t0}
	\mathrm{Re}\, v= g_s\mathrm{Re}\, v_1 + \cO(g_s^3) \, ,\qquad \mathrm{Re}\, v_1=\frac12 e^{A_0} d f_\star \,,\qquad  f_\star \equiv -\frac5m \left(3A_2 - \delta\phi_2 - \frac12 \psi_1^2\right)\,.
\end{equation}
As for $j$ and $\omega$, we want them to reconstruct in the limit an $SU(3)$-structure $(J,\Omega)$. We will simply assume here the latter to be fixed, and $(j,\omega)$ to be determined by (\ref{eq:JO-jo}). 
We don't know whether this assumption is really warranted at higher orders in our expansion, but up to the order of our computations we will see no difference. All this leads to 
\begin{equation}\label{eq:lim-JO}
	J_\psi = J + \cO(g_s^2)  \, ,\qquad \Omega= \frac i{\psi_1} v_1 \wedge \omega_0 + \cO(g_s^2)\,. 
\end{equation}
In fact we will be able to write everything in terms of $v$ and the fixed $(J, \Omega)$. 
It should be remarked that we are not assuming these to be those of the underlying Calabi--Yau, since we are aiming at removing the smearing. 

$\Omega$ still defines an \emph{almost} complex structure $I$: we mentioned in section \ref{sub:10d-susy} that it is at every point the wedge product of three one-forms $h^a$, which are then defined to be the $(1,0)$-forms of $I$. In fact we see from (\ref{eq:lim-JO}) that $v_1$ is one of these $(1,0)$-forms, and we can use this to determine $\mathrm{Im}\, v_1$ in the expansion $\mathrm{Im}\, v\sim g_s \mathrm{Im}\, v_1 + O(g_s^3)$.  But we are not assuming $I$ to be integrable; this would be implied by $d \Omega=0$, which is not part of the equations we found in section \ref{ssub:t0}. On the other hand, at leading order (\ref{eq:dJpsit0}) simply becomes 
\begin{equation}\label{eq:dJ0}
	dJ=0\,.
\end{equation}

The metric is not really needed to find a solution; it is determined by the forms of the $SU(3)\times SU(3)$-structure. The procedure comes originally from generalised complex geometry \cite{gualtieri}, and was explained in detail in \cite[Sec.~2.2.2]{Saracco:2012wc}. Fortunately at the leading order we are working with, the procedure reduces to the simpler one for $SU(3)$-structures, which we will illustrate in an example later on. In terms of this metric, one can invert the relation for $\Omega$ in (\ref{eq:lim-JO}) with a contraction:
\begin{equation}
	\omega_0=-\frac i{2 \psi_1} \bar v_1\cdot \Omega\,. 
\end{equation}

One last comment about the geometric forms: we are taking the volume of the internal space $X_6$ to be $\mathrm{Vol}(X_6)\sim R^6$, but we are taking care of that by scaling coordinates rather than the metric and forms. So for example for the torus cases below, the periodicities of the internal coordinates will scale like
\begin{equation}
	\Delta y \sim R\,.
\end{equation}
One can of course easily always switch to another point of view, where the coordinates don't rescale and forms do; this would lead to $J\sim R^2 J_0$, $\Omega\sim R^3 \Omega_0$. We take this viewpoint in the explicit example of section \ref{ss:torus}.

Applying the above procedure to \eqref{eq:flux-t0} we obtain the following relations between the fluxes and the $SU(3) \times SU(3)$-structure forms:
\begin{subequations}
\label{eq:leading-fluxes}
\begin{align}
	\label{Hleading}
	\hat H&= \frac25 F_0 g_s e^{-A_0} \mathrm{Re}\, \Omega + \cO(g_s^3)\,,\\
	\label{F0leading}
	F_0 &=  F_0 e^{-4A_0} - J\cdot d(e^{-3A_0}\mathrm{Im}\, v_1) + \cO(g_s^2) \,,\\
	\label{F2leading}
	F_2 &= -\frac{1}{g_s}  J \cdot d(e^{-3A_0} \mathrm{Im}\, \Omega) + \cO(g_s)\,,\\
	\label{F4leading}
	 F_4 &= F_0 J^2 \left(\frac12 - \frac{1}{5} e^{-4A_0}\right) +  J \wedge d(e^{-3A_0} \mathrm{Im}\, v_1) + \cO(g_s^2)\,,\\
	 F_6 &= 0\,.
\end{align}
\end{subequations}
Where recall that $F_0 = \ell_s^{-1} m$. Using (\ref{eq:GF-t0}) we find the physical fluxes
\begin{equation}
	H= \hat H + g_s d(\psi_1 \mathrm{Im}\, \omega_0) + \cO(g_s^3)\, ,\qquad
	\mathbf{G} = e^{ (g_s\psi_1 \mathrm{Im}\, \omega_0 + \cO(g_s^3))\wedge} \mathbf{F}\,.
\end{equation}
Notice that $(\mathrm{Im}\, \omega_0)^3=0$, so the exponential truncates. 

To summarise, in order to find a solution in this limit we need to find an $SU(3)$-structure $(J,\Omega)$, a $(1,0)$-form $v_1$, and a funcntion $A_0$, such that $e^{A_0}{\rm Re}\, v_1$ is exact, $J$ is closed  ((\ref{eq:Rev1-t0}),  (\ref{eq:dJ0})). When plugged into (\ref{eq:leading-fluxes}) these should provide an expression for the fluxes that solves the Bianchi identities, up the order of the approximation.


\section{Solving the Bianchi identities}
\label{sec:Bianchi}

In this section we solve exactly the Bianchi identities for the internal sources that correspond to \cite{DeWolfe:2005uu}. For this we consider a manifold $X_6$ that admits a Calabi--Yau metric, namely a metric of $SU(3)$ holonomy, so that we can have a 10d interpretation of the sources that appear in \cite{DeWolfe:2005uu}. Looking at the first relation in \eqref{intflux} and the expression for $H$ in terms of $SU(3)\times SU(3)$ structures with $\theta=0$ one infers that it must be of the form
\be
H =  2\mu \re \Omega_{\rm CY} + d\tilde{B}\, ,
\label{Hini}
\ee
with $5 \mu  \langle g_s^{-1}\rangle= G_0$.  Let us for now set $\tilde{B} = 0$ and solve the Bianchi identities in this case, and then recover the general solution by applying a $B$-field transformation. For the particular solution we denote the RR fluxes by $\tilde{F}_{2p}$. 

The Bianchi identity for the two-form flux reads
\be
  \ell_s^2 d\tilde{F}_2 =  2 m \hat{\mu} \re \Omega_{\rm CY} -4  \d(\Pi_{\rm O6}) \, , 
  \label{BIF2ini}
\ee
with $\hat{\mu} = \mu\ell_s$. By Hodge decomposition the most general solution is of the form
\be
 \tilde{F}_2 =   d^\dag_{\rm CY} K + \tilde{F}_2^{\rm h}  + dC_1\, ,
\label{F2K}
\ee
with $dC_1$ exact, $\tilde{F}_2^{\rm h}$ Calabi--Yau harmonic, and $d^\dag_{\rm CY}$ constructed with the Calabi--Yau metric. 
Finally, $K$ is a 3-form current that always exists, as it satisfies the following Laplace  equation
\be
\ell_s^2 \Delta_{\rm CY} K =  2m\hat{\mu}\re \Omega_{\rm CY} -4 \d(\Pi_{\rm O6})\, ,
\label{defK}
\ee
where $\Delta_{\rm CY} = d^\dag_{\rm CY} d + d d^\dag_{\rm CY}$ is constructed from the CY metric. Indeed, following  \cite[sec.~3.4]{Hitchin:1999fh} notice that $d\re \Omega_{\rm CY} = d \, \d(\Pi_{\rm O6}) = 0$, $\Delta_{\rm CY} d K =0$ and $dK$ is harmonic. Because it is also exact, then necessarily $dK=0$. We conclude that $\Delta_{\rm CY} K = dd^\dag_{\rm CY} K$ from where \eqref{F2K} follows.  One can then constrain $K$ by using that $J_{\rm CY}$, $\Omega_{\rm CY}$ are covariantly constant with respect to $\Delta_{\rm CY}$: 
\begin{align}
\Delta_{\rm CY} K \wedge J_{\rm CY} = 0 &  \Rightarrow  \Delta_{\rm CY}(K \wedge J_{\rm CY}) = 0 \ \Rightarrow \  K \wedge J_{\rm CY} = 0\, , \\
\Delta_{\rm CY} K \wedge \re \Om_{\rm CY} = 0  & \Rightarrow  \Delta_{\rm CY}(K \wedge \re \Om_{\rm CY}) = 0  \Rightarrow   K = \varphi \re \Om_{\rm CY} + c \im \Om_{\rm CY} + \re k ,
\label{formK}
\end{align}
with $\varphi$ a real function, $c$ a constant that we will take to be zero, and $k$ a (2,1) primitive current. Here we have used that there are no harmonic 5-forms in the CY metric. One then obtains that
\be
\label{F2H}
 d^\dag_{\rm CY} K  =   \star_{\rm CY} \left( d\varphi \wedge \im  \Om_{\rm CY} \right) -  \star_{\rm CY} d \im   k  = - J_{\rm CY} \cdot d \left(2 \varphi \im  \Om_{\rm CY} \right) - V_2  \, ,
\ee
where $V_2$ is a primitive (1,1)-form in the CY sense. One can check that this implies that
\be
\tilde{F}_2 = - J_{\rm CY} \cdot d( 4 \varphi \im \Om_{\rm CY} - \star_{\rm CY} K)  + \tilde{F}_2^{\rm h}  + dC_1 \, .
\label{keyimOm}
\ee

As for the remaining fluxes, it is easy to see that
\be
\tilde{F}_4 =   \tilde{F}_4^{\rm h} - 4 \mu \varphi \, J_{\rm CY} \wedge J_{\rm CY} +  2\mu \re \Omega_{\rm CY} \wedge C_1 + dC_3  \, ,
\label{F4H}
\ee
with $dC_3$ exact, $\tilde{F}_4^{\rm h}$ Calabi--Yau harmonic, satisfies the Bianchi identity $d\tilde{F}_4 = 2\mu \re \Omega_{\rm CY} \wedge \tilde{F}_2$. As for the six-form flux, we can set $\tilde{F}_6 = dC_5$ to be an exact form. 

Finally, whenever $\tilde{B}$ in \eqref{Hini} is not trivial, the solution for the fluxes will be given by
\be
\mathbf{G}= e^{\tilde{B} \wedge} \mathbf{\tilde{F}}\, ,
\ee
with $\tilde{F}_0 = G_0$ and the remaining $\tilde{F}_{2p}$ as specified.

\section{Solving the supersymmetry equations}
\label{sec:solution}

Thanks to our previous results, in this section we will be able to give a 10d supersymmetric background describing the  DGKT relations \eqref{intflux} for any manifold $X_6$ that admits a Calabi--Yau metric. Our strategy will be simple: we will  provide expressions for $\Om$, $J$, $e^{A_0}$ and $v_1$ in terms of Calabi--Yau quantities, such that when plugged in \eqref{eq:leading-fluxes} provide backgrounds fluxes solving the Bianchi identities up to the appropriate order of the expansion. Because of that, our background can only be thought of as an approximation to an actual supersymmetric solution describing a 10d counterpart of \cite{DeWolfe:2005uu}. This approximation becomes more accurate in the limit of large volume and weak coupling, approaching the $SU(3)$-structure smeared solution in that limit.

\subsection{General Calabi--Yau manifolds}

Since by assumption $X_6$ admits a Calabi--Yau metric, we can profit from the discussion in section \ref{sec:Bianchi} as a guiding principle to construct the $SU(3)\times SU(3)$-structure metric in $X_6$. First, as the two-form $J$ is closed, we will assume that
\be \label{eq:J0=JCY}
J = J_{\rm CY} +  \cO(g_s^2)\, ,
\ee
where recall that $g_s = 5\mu/m = 5 \cV_{X_6}^{-1/2} / m$. Then, one can guess the form of $\im \Om$ by comparing \eqref{F2leading} and \eqref{keyimOm}. Indeed, let us consider the following expression
\be 
e^{-3A_0} \im \Om = \left(1 +  g_s 4 \varphi \right)  \im \Omega_{\rm CY} - g_s \star_{\rm CY} K + \cO(g_s^2)\, ,
\label{AimOm}
\ee
with $K$ an exact three-form defined by \eqref{defK} and $\varphi$ defined by \eqref{formK}. Plugging this into \eqref{F2leading} one obtains \eqref{keyimOm} with $\tilde{F}_2^{\rm h}= dC_1 =0$. Therefore with the choice \eqref{AimOm}, $F_2$ in \eqref{F2leading} satisfies  \eqref{BIF2ini}. Finally, such an $F_2$ also satisfies the Bianchi identity up to $\cO(g_s^2)$ terms  if we assume that $\tilde{B} \sim \cO(g_s^2)$ in \eqref{Hini}, as we will do in the following.

From here, one may construct the rest of $\Omega$. In general its real and imaginary parts are related by a method in \cite{Hitchin:2000jd} and reviewed in \cite[Sec.~3.1]{Tomasiello:2007zq}. In particular here we can consider ${\rm Im}\,\Omega$ as a perturbation over ${\rm Im}\,\Omega_{\rm CY}$, and apply the perturbation formulas \cite[(3.8)--(3.10)]{Tomasiello:2007zq}. One finds that
\be
e^{-3A_0} \re \Om  =\left( 1 + g_s 2\varphi \right) \re \Om_{\rm CY} + g_s K +\cO(g_s^2)\, ,
\ee
which satisfies the $SU(3)$-structure relation $\re \Om \wedge \im \Om = \frac{2}{3} J^3$ provided that
\be 
e^{-4A_0} =  1 + g_s 4 \varphi + \cO(g_s^2)\, .
\label{Awarp}
\ee
From the definition of $\varphi$ it is easy to see that $\int_{X_6} \varphi = 0$ and therefore $\langle e^{-4A} \rangle = 1$ up to this order of approximation, as expected. This moreover leads to
\be
e^{-A_0} \re \Om  = \re \Om_{\rm CY} + g_s K + \cO(g_s^2)\, ,
\label{AreOm}
\ee
and so, since $dK=0$, the Bianchi identity $dH = 0$ is satisfied up to order $\cO(g_s^3)$. Finally, when plugging \eqref{AreOm} into \eqref{Hleading} we obtain that $d\tilde{B} =  2\mu g_s K +  \cO(g_s^2)$, consistently with our assumption. We finally obtain
\bea \label{eq:Omega0-def}
\Omega & = & 
\Omega_{\rm CY}  + g_s  k + \cO(g_s^2) \, ,
\eea
where $k$ the primitive  (2,1)-from $k$ defined by \eqref{formK}. Notice that this expression is compatible with  SU(3)-structure torsion classes, since
\be
d\Om = g_s dk = - d\varphi \wedge \Om_{\rm CY} + i g_s V_2 \wedge J = dA_0 \wedge \im \Om + i W_2 \wedge J +\cO(g_s^2)\, ,
\ee
with $W_2 = g_s V_2 +\cO(g_s^2)$. So the leading correction to the pair $(J, \Omega)$ corresponds to a non-Ricci-flat, symplectic metric with SU(3)-structure.

With this choice of $(J, \Om)$ and warp factor it is easy to accommodate the remaining expressions in \eqref{eq:leading-fluxes} to satisfy the Bianchi identities with $\tilde{B} \sim  \cO(g_s^2)$. Indeed, to fit \eqref{F4leading} into \eqref{F4H} one simply needs to take
\be \label{eq:C3-pert}
\tilde{F}_4^{\rm h} = \frac{3}{10} F_0\, J_{\rm CY} \wedge J_{\rm CY} \, , \qquad C_3 = e^{-3A} J_{\rm CY}\wedge \im v_1 \, .
\ee
Finally, the Bianchi identity for $F_0$ is compatible with the rhs of \eqref{F0leading} if one takes $\im v_1$ to be the imaginary completion of 
\be \label{eq:v1pert}
\re v_1 = \oh e^{A_0} d f_\star \, ,
\ee
with
\be
\ell_s \Delta_{\rm CY} f_\star  = - g_s m 8 \varphi  +  \cO(g_s^2) \, .
\label{defstar}
\ee
In other words, $\im v_1 = I\cdot \re v_1$, with $I\cdot$ the action of the complex structure. At the leading order of the  expansion, this implies that $v_1 =  \p_{\rm CY} f_\star$. It follows from this result that the $B$-field transformation in \eqref{eq:GF-t0} is suppressed by $g_s^2$ and does not induce any change in the fluxes at the present order.
In the following we will discuss in detail how this approximate solution looks like in the case of a toroidal orbifold, where the above expressions can be made more explicit.

In summary, our solution is specified by the $SU(3)$-structure given in \eqref{eq:J0=JCY}, \eqref{eq:Omega0-def}; the one-form $v_1$ specified by \eqref{eq:v1pert}; and the warping function $A_0$ in \eqref{Awarp}. The Bianchi identity were shown to be solved for $F_2$ in \eqref{AimOm}; for $H$ in \eqref{eq:Omega0-def}; for $F_0$ in \eqref{defstar}; for $F_4$ in \eqref{eq:C3-pert}. By the general results of \cite{Koerber:2007hd}, once the supersymmetry equations and Bianchi identities are satisfied, the equations of motion for the fields are also solved.

\subsection{A toroidal orbifold example}
\label{ss:torus}

Let us consider the particular case where $X_6 = T^6/\IZ_2\times\IZ_2$, as in \cite{Camara:2005dc}. We consider the choice of discrete torsion that makes it T-dual to the closed string background in \cite{Berkooz:1996km}, so that all O6-planes have negative charge and tension. In the orbifold limit, the Calabi--Yau structure is essentially inherited from the covering space $T^6$, so we can write
\bea
J_{\rm CY} & = & 4\pi^2 t_i dx^i \wedge dy^i \, ,\\
\re \Om_{\rm CY} & = &  h\left(\tau_1\tau_2\tau_3 \b^0 - \tau_1 \b^1 - \tau_2 \b^2 - \tau_3 \b^3 \right) \, ,\\
\im \Om_{\rm CY} & = &   h\left( \a_0 - \tau_2\tau_3 \a_1 - \tau_1\tau_3 \a_2 - \tau_1\tau_2 \a_3 \right) \, ,
\eea
where
\be
t^i  =  R_{x^i} R_{y^i}\, , \qquad \tau_i = \frac{R_{y^i}}{R_{x^i}}\, , \qquad h =  8\pi^3 \sqrt{\frac{t_1 t_2 t_3}{\tau_1\tau_2\tau_3}} =  8\pi^3 R_{x^1}R_{x^2}R_{x^3 }\, ,
\ee
and we have the following basis of bulk three-forms
\bea\nonumber
\a_0 = dx^1 \wedge dx^2 \wedge dx^3\, , & \quad & \b^0 = dy^1 \wedge dy^2 \wedge dy^3 \, ,\\ \nonumber
\a_1 = dx^1 \wedge dy^2 \wedge dy^3\, , & \quad & \b^1 = dy^1 \wedge dx^2 \wedge dx^3 \, ,\\ \nonumber
\a_2 = dy^1 \wedge dx^2 \wedge dy^3\, , & \quad & \b^2 = dx^1 \wedge dy^2 \wedge dx^3 \, ,\\ \nonumber
\a_3 = dy^1 \wedge dy^2 \wedge dx^3\, , & \quad & \b^3 = dx^1 \wedge dx^2 \wedge dy^3 \, .
\eea
In principle one can consider partially cancelling the charge of the O6-planes with D6-branes on top of them, and so different choices of $H$-flux that will cancel the corresponding generalisation of the tadpole condition \eqref{tadpole}. For simplicity, we will consider those cases where the $H$-flux is of the form 
\be
\ell_s [H] = 8 q\left([\b^0] - [\b^1] - [\b^2] -[\b^3] \right) \, ,
\ee
for some choice of $q \in \IZ$, with $q m = 4$ in the particular case where no D6-branes are present. Then supersymmetry requires that
\be
\tau_1=\tau_2=\tau_3 = 1\, , \quad \text{and} \quad \hat{\mu} = \frac{4q}{h}\, ,
\ee
from where it is clear that $\hat{\mu} \sim \cV^{-1/2}_{X_6}$. In this setup we find a solution for \eqref{defK} of the form
\be
K =  qm  \left(B_0  \b^0 - B_1  \b^1 - B_2 \b^2 - B_3  \b^3 \right)\, ,
\ee
with
\bea
\label{Bs}
B_0 & = & - h^{2/3}  \sum_{\vec{\eta}} \sum_{\vec{0}\neq\vec{n}\in \IZ^3} \frac{e^{2\pi i  \vec{n}\cdot \left[(y^1, y^2,y^3)+\vec\eta\right]}}{4\pi^2\vec{n}^2} \, , \
 B_1  =  - h^{2/3} \sum_{\vec{\eta}} \sum_{\vec{0}\neq\vec{n}\in \IZ^3} \frac{e^{2\pi i  \vec{n}\cdot \left[(y^1, x^2,x^3)+\vec\eta\right]}}{4\pi^2\vec{n}^2} \, , \qquad   \\ \nonumber 
 B_2 & = & -  h^{2/3} \sum_{\vec{\eta}}  \sum_{\vec{0}\neq\vec{n}\in \IZ^3} \frac{e^{2\pi i  \vec{n}\cdot \left[(x^1, y^2,x^3)+\vec\eta\right]}}{4\pi^2\vec{n}^2} \, , \
 B_3  =  - h^{2/3} \sum_{\vec{\eta}} \sum_{\vec{0}\neq\vec{n}\in \IZ^3} \frac{e^{2\pi i  \vec{n}\cdot \left[(x^1, x^2,y^3)+\vec{\eta}\right]}}{4\pi^2\vec{n}^2} \, , \qquad 
\eea
where for simplicity we have set $R_{y^1} = R_{y^2} = R_{y^3} = R$,\footnote{Otherwise one should replace $\vec{n}^2/R^2$ by $|\vec{n}|^2 = \sum_i \left(n_i/R_{y^i}\right)^2$.} and $\vec\eta$ has entries which are either 0 or $1/2$. Notice that $d(B_i\b^i) = 0\, \forall i$ so that $K$ is closed, and in fact exact.

It is important to point out that the expansion for the $B_i$'s in terms of Fourier modes should be understood as a formal solution since the sum is not convergent. A regularised version of these Green functions using the Jacobi theta function was originally suggested in \cite{Shandera:2003gx} and some details have been recently studied in \cite{Andriot:2019hay}. For practical purposes, the regularised functions behave as standard Green functions in flat space when approximating the source and go to zero as we move away. 

Following the Calabi--Yau general discussion, can rewrite things as \eqref{formK} with 
\be
\varphi  =  \frac{qm}{4h}  \sum_{i=0}^3 B_i \, ,\qquad \re k   =  K - \varphi \re \Om^{\rm CY} \, .
\ee
Notice that $\varphi \sim \cO(R^{-1})$ but it is not suppressed by an extra factor of $g_s$. Eq.\eqref{AimOm} becomes 
\be
e^{-3A_0} \im \Om = C_0 \a_0 - C_1 \a_1 - C_2 \a_2 - C_3 \a_3 + \cO(g_s^2)\, ,
\label{AimOmT}
\ee
with $C_i = h - g_s q m \left(B_i - \sum B_i\right)$. Stability techniques from \cite{Hitchin:2000jd} (reviewed for example in  \cite[Sec.~3.1]{Tomasiello:2007zq}) tell us when a three-form can be the imaginary part of a decomposable form $e^{-3A_0}\Omega$, and what the real part is. For \eqref{AimOmT} this tells us that $\Omega$  exists in regions where $C_0 C_1 C_2 C_3 > 0$, and determines
\bea
e^{-A_0} \re \Om &= & h^{2/3}  \left(C_0C_1C_2C_3\right)^{1/3} \left[C_0^{-1} \b^0  - C_1^{-1} \b^1  -C_2^{-1} \b^2  -C_3^{-1} \b^3   \right] + \cO(g_s^2)  \\ \nonumber
& = & \re \Om_{\rm CY} +  g_s K + \cO(g_s^2)\, .
\eea
where we have used that from imposing the relation $\re \Om \wedge \im \Om = \frac{2}{3} J_{\rm CY}^3$ one obtains
\be
e^{-4A_0} = h^{-4/3} \left(C_0C_1C_2C_3\right)^{1/3} =  1 + 4g_s \varphi +  \cO(g_s^2)\, ,
\ee
in agreement with \eqref{Awarp} and \eqref{AreOm}. By combining all these expressions we obtain
\be
\Om = i  e^{3A_0} C_0  \left[ dx^1 + i  \left(\frac{C_2C_3}{C_0C_1}\right)^{1/2}\hspace{-.3cm}  dy^1 \right] \wedge  \left[ dx^2 + i  \left(\frac{C_1C_3}{C_0C_2}\right)^{1/2}\hspace{-.3cm}  dy^2 \right] \wedge  \left[ dx^3 + i  \left(\frac{C_1C_2}{C_0C_3}\right)^{1/2}\hspace{-.3cm}  dy^3 \right] \, ,
\ee
which corresponds to the metric
\bea
ds^2 & = & h^{2/3} \left[\left(\frac{C_0C_1}{C_2C_3}\right)^{1/2}(dx^1)^2 + \left(\frac{C_0C_2}{C_1C_3}\right)^{1/2}(dx^2)^2  + \left(\frac{C_0C_3}{C_1C_2}\right)^{1/2}(dx^3)^2\right] \\ \nonumber
& + &  h^{2/3} \left[\left(\frac{C_2C_3}{C_0C_1}\right)^{1/2}(dy^1)^2 + \left(\frac{C_1C_3}{C_0C_2}\right)^{1/2}(dy^2)^2  + \left(\frac{C_1C_2}{C_0C_3}\right)^{1/2}(dy^3)^2\right] + \cO(g_s^2)\, .
\eea
Regarding the fluxes, we find that
\bea
H & = & \frac{2}{5}F_0 g_s \left(\re \Om_{\rm CY} + g_s K \right) - g_s\oh   d\re \left(\bar{v}_1 \cdot \Omega_{\rm CY} \right) + \cO(g_s^3)  \, , \\
 F_2 & = &  d^{\dag}_{\rm CY} K  + \cO(g_s)  \, , \\
F_4 & = & F_0 J_{\rm CY}^2 \left(\frac{3}{10}  - \frac{4}{5} g_s \varphi \right)  + J_{\rm CY} \wedge d \im v_1 + \cO(g_s^2) \, ,
\eea
where $\im v_1$ is the imaginary completion of
\be
2 e^{-A_0} \re v_1  =  df_\star + \cO(g_s^2) \, , \qquad {\rm with} \qquad 
f_\star  =  - \ell_s g_s \frac{2qm^2}{h} \sum_i \tilde{B}_i\, .
\ee
where $\tilde{B}_i$ stand for the functions $B_i$ in \eqref{Bs} with the replacement $R^2/\vec{n}^2 \rightarrow R^4/|\vec{n}|^4$. Note that, unlike the $B_i$, the $\tilde{B}_i$ can be shown to be convergent, so there is no need to regularise them.

\subsection{Comparison with the smeared solution}
\label{ss:comparison}

Let us summarise our approximate solution. We obtain that the background fluxes are given by $\ell_s G_0 = m$ and
\begin{subequations}
	\label{solutionflux}
\begin{align}\nonumber
H & =  g_s \frac{2}{5}G_0 \left(\re \Om_{\rm CY} + g_s K \right) - \oh   d\re \left(\bar{v} \cdot \Omega_{\rm CY} \right) + \cO(g_s^{3})= g_s \frac{2}{5}G_0 \re \Om_{\rm CY}  \left( 1+ \cO(g_s^{4/3})\right) \, , \\
G_2 & =   d^{\dag}_{\rm CY} K  + \cO(g_s)  = \cO( g_s^{2/3}) \, , \\
G_4 & =  G_0 J_{\rm CY}^2 \left(\frac{3}{10}  - \frac{4}{5} g_s \varphi \right) + J_{\rm CY} \wedge g_s^{-1}d \im v  + \cO(g_s^2) = \frac{3}{10} G_0 J_{\rm CY}^2 \left( 1+ \cO(g_s^{4/3})\right)\, , \\
G_6 & = 0\, ,
\end{align}
\end{subequations}   
where  $g_s = 5 \cV^{-1/2}_{X_6}/m$, $K$ is defined by \eqref{defK} and the proof of $G_6=\left( e^{b}\bf F\right)|_6= 0$ is given in appendix \ref{sub:teq0}. The warp factor, dilaton and internal metic are specified by
\begin{subequations}	
	\label{solutionsu3}
\begin{align}
e^{-A}  & = 1 + g_s \varphi + \cO(g_s^2) \,  =\, 1+ \cO(g_s^{4/3}) \, ,\\
e^{\phi}  & = g_s \left(1 - 3  g_s \varphi\right) + \cO(g_s^3) \, =\, g_s \left( 1 + \cO(g_s^{4/3}) \right) \, ,\\
\Om & = \Om_{\rm CY} + g_s k +  \cO(g_s^2) \, =\, \Om_{\rm CY} \left( 1 + \cO(g_s^{4/3}) \right)\, , \\
J & = J_{\rm CY} + \cO(g_s^2) \, =\, J_{\rm CY} \left( 1 + \cO(g_s^{4/3}) \right)\, , \\
v & = g_s \p_{\rm CY} f_\star + \cO(g_s^3) \, =\, \cO(g_s^{2})
\end{align}
\end{subequations}     
where recall that $\varphi$ and $k$ are defined by \eqref{formK}, and $f_\star$ by \eqref{defstar}. When next to a $p$-form, the above scalings $\cO(g_s^k)$  are to be interpreted with respect to the natural scaling of the  $p$-form, so the total scaling of the object is $\cO(g_s^{k-p/3})$. 

We notice that the natural parameter of the expansion is $g_s^{4/3} \sim \cV_{\rm CY}^{-2/3}$, or in other words the quantum of four-form flux $G_4$.  We also notice that at leading order we recover precisely the Calabi--Yau background with fluxes \eqref{eq:SU3th0}. At next order our solution is an $SU(3) \times SU(3)$ background, which contains an SU(3)-structure pair $(J, \Omega)$ with the following torsion classes 
\bea
\label{Omsu3}
d\Omega & = & i W_2 \wedge J + d(\phi-2A) \wedge \Omega\\
dJ & = & 0
\eea
and with $e^{\phi-3A} = g_s$. While this is the starting point for the analysis of SU(3)-structure backgrounds with $\theta=0$, the difference here is that a varying warp factor is allowed. This is thanks to the presence of a non-trivial one-form $v$.
This varying warp factor, and in general the three-form $K$ obtained from solving the Bianchi identity for $G_2$ at leading order, also modifies the fluxes $H$ and $G_4$ at this order, adding a non-CY-harmonic piece. 

In view of the no-go results for SU(3)-structure compactifications of section \ref{sec:nogosol}, one may wonder how this approximate background at $\cO(g_s)$ can overcome the obstructions therein. In particular let us see how \eqref{tf2loc} is modified to allow for a non-smeared solution. First notice that to arrive to this equation one uses that \cite{Lust:2004ig}
\be
dG_2 \wedge \Omega + G_2 \wedge d \Omega = d (G_2 \wedge \Omega)\, .
\ee
In type IIA SU(3)-structure compactifications $G_2 \wedge \Omega \equiv 0$ iff the warp factor is constant, as one can show from \eqref{su3fluxG2} and the general expression for $d\Omega$. In the case of a SU(3)-structure background this 
follows when we impose that the rhs of \eqref{su3fluxG0} is closed. In our more general background this expression generalises to \eqref{F0leading} allowing for a non-constant, subleading piece or the warp factor, as the solution shows explicitly. This in turn implies that $d(G_2 \wedge \Omega) \neq 0$ adding the extra term to \eqref{tf2loc}. In our solution this term is comparable to the terms in the lhs of \eqref{tf2loc} wedged with $\im \Omega$, which would then scale like $\Om$. Therefore the cancellation of this term is possible away from the localised sources and no smearing is needed. 

Notice that this the term $d (G_2 \wedge \Omega)$ is a non-trivial contribution to $X_5$ in the source balanced equation \eqref{preGOE}. So let us analyse how this more general equation can be satisfied for our approximate solution. Using the background in \eqref{solutionflux} and \eqref{solutionsu3} one obtains
\begin{subequations}
	\label{balancedsol}
\begin{align}
 \label{dX5sol1}
3\mu^2 e^{-4A} \left< \mathrm{Re\;}\Phi_+ , \mathrm{Im\;}\Phi_+ \right>  \sim  -\frac{2}{25} G_0^2J^3  & \sim   \cO(g_s^0)\, d\mathrm{vol}_{X_6}\, , \\
 \label{dX5sol2}
 e^{4A}  \sum\nolimits_k G_k \wedge \star G_k  \sim \frac{26}{75}G_0^2J^3+\frac{ |G_2|^2}{6}J^3 & \sim   \cO(g_s^0)\,  d\mathrm{vol}_{X_6}+ \cO(g_s^{4/3})\,  d\mathrm{vol}_{X_6} \, ,\\
 -dX_5  \sim -\frac{4}{15}G_0^2 J^3-\delta^{(3)}_{\mathrm{source}}\wedge \frac{1}{g_s}\im\Omega_{\text{CY}} & \sim \cO(g_s^0)\,  d\mathrm{vol}_{X_6}\, ,
 \label{dX5sol}
\end{align}
\end{subequations} 
where, although $G_2\wedge \star G_2\sim \cO(g_s^{4/3})$ at leading order, we are writing explicitly this term to make easier the comparison with \eqref{tf2loc}.  Even if the sum of the first two terms gives a positive definite quantity --- recovering the case $\theta=0$ in \eqref{tf2loc} --- the terms coming from $dX_5$ are able to compensate this contribution. Indeed, for the case at hand one can check that the leading contribution to \eqref{dX5sol} comes from $d\left(G_2 \wedge \left( e^A \mathrm{Im\;}\Phi_- \right)_{3}\right)$ and that it cancels the other two contributions at order $\cO(g_s^0)$.
 In fact, one can easily check that this corresponds to the contribution $d (e^{-\phi} G_2 \wedge \Omega)$ that would allows to circumvent the obstruction related to \eqref{tf2loc}.

\section{Conclusions}
\label{sec:conclu}

In this paper we have found approximate solutions to the ten-dimensional supersymmetry equations which exhibit some key features of the DGKT four-dimensional vacua \cite{DeWolfe:2005uu}. The solutions are first order in an expansion parameter  corresponding to the average 10d dilaton $g_s$, or equivalently to the AdS$_4$ scale $\mu$ or $\cV_{X_6}^{-1/2}$ in string units.

The solutions are such that in the limit $g_s\rightarrow 0$ the background metric of the Calabi--Yau $X_6$ is the Ricci-flat one and the warp factor is constant. The non-vanishing fluxes are $G_0$ and $G_4 = \frac{3}{10} G_0 J_{\rm CY}^2$. This background corresponds to the smeared-O6-plane solution to DGKT proposed in \cite{Acharya:2006ne}. For small but non vanishing $g_s$, corrections to this background appear. The leading ones can be described in terms of the solution to the Bianchi identity $dG_2 = G_0H + \d_{\rm O6}$, which defines a function $\varphi$ and a (2,1)-form $k$. The first one corrects the warp factor and the dilaton, and the second one the three-form $\Omega$. Due to this metric deformation $X_6$ becomes a manifold with $SU(3)\times SU(3)$-structure.\footnote{This can also be thought of as $SU(3)$-structure with an additional 1-form. The $SU(3)$-structure part has the same torsion classes as type IIA Minkowski backgrounds with O6-planes.} Finally, $H$ and $G_4$ are also corrected in terms of $\varphi$ and $k$, no longer being harmonic forms in the Calabi--Yau sense.

Given that our solution was obtained in an expansion in the average string coupling $g_s$, one might wonder whether it competes with the genus expansion in string theory. Since we are at weak coupling, certainly the leading order part of the solution is under good control. The next order comes in at $g_s^{4/3}$. We expect that string loop corrections should appear at order $g_s^2$ or higher, and therefore the analysis should hold at least to first order. We leave a more detailed analysis of the magnitude of string corrections in this background for future work. 

Perhaps an even more delicate issue is the fact that we have solved the equations at the two-derivative level, so at leading order in $\alpha'$. Higher $\alpha'$ corrections are controlled by the curvature radius which is again related to our expansion parameter $g_s \sim R^{-3}$. In this case a more accurate analysis of the magnitude and, importantly, the precise form of such corrections is needed to see whether their effect is substantial. 

Regarding the issue of scale separation and the Strong ADC, our results show that the DGKT proposal for scale separation has passed a first non-trivial test. There could have been an obstruction manifest already at first order in the supersymmetry equations, but we have shown that this is not the case (at least at the two-derivative level). 

However, they are still far from settling the issue. Before even asking about separation of scales we may ask whether a full ten-dimensional solution actually exists. At a technical level, a first possible obstruction may appear at the next order in the expansion parameter. Indeed, a crucial part of DGKT is that it involves intersecting sources, and ten-dimensional solutions of such sources are poorly understood. At the first, linearised, level of the expansion the interactions between the sources drop out which is why we are able to find a solution relatively easily. The interactions only appear at the next level, where this feature of the construction is first tested. Note that if a solution does exist, it would be interesting to see if it also realises the picture proposed in \cite[Sec.~5.3]{Marchesano:2019hfb} in which the pure spinors $\Phi_{\pm}$ differ from the Calabi--Yau ones only by non-harmonic forms.

There are also other, older and more general, open problems with any solutions of massive type IIA with O-planes. In our approximate solution, near an O6-plane we obtain an   warping behaviour of the form $e^{-4A} \sim 1-\frac{g_s\ell_s}{r}$ \cite{Andriot:2019hay}.  This defines a region in which we enter strong coupling and the supergravity approximation breaks down.
Typically the O-plane singularities may be resolved by uplifting to M-theory or F-theory. This is not possible here due to the mass parameter, and so the fate of these singularities remains an open question. It should be noted that the mass parameter, which obstructs an M-theory uplift \cite{Aharony:2010af}, is the crucial element to obtaining scale-separated vacua (it cannot be turned off, unlike some of the other fluxes). Practically, what this means is that we are simply not able to say anything about what happens near the O-planes in our solution. The hope is therefore that either we make progress on understanding the O-planes, or that we are able to settle the relevant questions without needing to worry about them.\footnote{An interesting possibility is that the singularities are removed already in the IIA supergravity description, similarly to the ideas in \cite{Saracco:2012wc}.}

There are other open questions, some of which were raised already in \cite{DeWolfe:2005uu}, such as the control of higher derivative terms in the presence of large fluxes parameters. Further, even if a ten-dimensional solution of string theory can be firmly proven under controlled approximations, it still remains to be checked that it really does exhibit separation of scales. For example, as we have shown, the solution exhibits dilaton gradients and warp factors which must be accounted for in establishing the mass scales of the KK and string modes.  

If the remaining open questions can be addressed and the property of scale separation proven, we would reach a significant result in string theory, and a counter example to the Strong ADC (the normal ADC is of course satisfied in DGKT). In such a scenario it would be interesting if there is a possible refinement of the Strong ADC which may hold. A particularly interesting proposal was made in \cite{Buratti:2020kda} related to the presence of discrete symmetries. In any case, we find it exciting and encouraging that the recent activity, and progress reported in this paper, suggests that at least the issue of DGKT and scale separation may be settled one way or the other in the not too distant future.

\bigskip 

\bigskip 

\vspace*{.05cm}

\centerline{\bf  Acknowledgments}

\vspace*{.05cm}

We would like to thank J.~Maldacena for collaboration in the initial stages of this project, and A.~Legramandi, A.~Uranga for useful discussions.  The work of F.M. and J.Q is supported by the Spanish Research Agency (Agencia Estatal de Investigaci\'on) through the grant IFT Centro de Excelencia Severo Ochoa SEV-2016-0597, and by the grant PGC2018-095976-B-C21 from MCIU/AEI/FEDER, UE. J.Q. is supported through the FPU grant No. FPU17/04293. A.T.~is supported in part by INFN and by the by MUR-PRIN contract 2017CC72MK003. We wish to thank the Simons Center for Geometry and Physics, Stony Brook University, where this work was initiated.

\appendix

\section{$SU(3)\times SU(3)$-structure compactifications}
\label{ap:SU33}

In this Appendix we will give more details about the pure spinor approach to supersymmetry.

\subsection{Types of pure spinor pairs}
As we mentioned in the main text, there are various solutions to the algebraic constraints that the pure forms $\Phi_\pm$ have to satisfy.  
\begin{itemize}
	\item $SU(3)$-structure. In this case the pure spinors are (\ref{eq:Phi-su3}), where $J$ is a real non-degenerate two-form, and $\Omega$ a complex three-form such that 
	\begin{equation}
		\Omega \wedge J= 0 \, ,\qquad  - \frac16 J^3 = -\frac i8 \Omega \wedge \bar \Omega \equiv \mathrm{vol}_6\,.
	\end{equation}
	Moreover, $\Omega$ should be \textit{decomposable}: at every point it should be possible to write it as the wedge product of three one-forms. This constraint allows to reconstruct ${\rm Im} \Omega$ from ${\rm Re} \Omega$, or viceversa; explicit formulas for this were given in \cite{Hitchin:2000jd} and reviewed for example in \cite[Sec.~3.1]{Tomasiello:2007zq}.
	
	\item $SU(2)$-structure. In this case, the data are those of a complex one-form $v$, a real two-form $j$ and a complex two-form $\omega$ which is again decomposable. They obey
	\begin{equation}
	\label{su(2)prop}
		j \wedge \omega = 0  \, ,\qquad  \omega \wedge \bar \omega = 2 j^2 \neq 0\,.
	\end{equation}
	Moreover, the contractions $v \cdot j = v \cdot \omega=0$. This latter condition can be regarded as part of the prescription on how to obtain the metric. The volume form is in this case $\mathrm{vol}_6 = -\frac i4 v \wedge \bar v \wedge j^2$.
	
	An equivalent definition is obtained by taking an $SU(3)$-structure $(J, \Omega)$ and adding a complex one-form $v$. The forms $j$ and $\omega$ are then obtained as
	\begin{equation}
		j= J - \frac i2 v \wedge \bar v \, ,\qquad \omega = \frac12 v \cdot \Omega\,.
	\end{equation}
	In other words, $\Omega= v \wedge \omega$.
		
	\item $SU(3)\times SU(3)$-structure. This is the generic solution. Here \cite[Sec.~2.2.2]{Saracco:2012wc}
	\begin{equation}
	\label{ap:su3phi}
		\Phi_+ = e^{3A-\phi} \cos\psi e^{i \theta } \exp[-i J_\psi] \, ,\qquad \Phi_- =e^{3A-\phi}\cos\psi v \wedge \exp[i \omega_\psi]\,,
	\end{equation}
	where $J_\psi$ and $\omega_\psi$ are defined in terms of an $SU(2)$-structure as in (\ref{eq:Jpsi}). 
\end{itemize}

\subsection{$SU(3)\times SU(3)$  $\theta \neq 0$} 
\label{sub:tneq0}

We now show the generic case, $\theta \neq 0$. The local solution in \cite{Saracco:2012wc} belongs to this class; however, we will see in the next section that this is unlikely to be promoted to a global solution.

Again there are first some purely geometric equations:
\begin{subequations}	
\begin{align}
	\label{eq:drho}
    &d(e^{3A-\phi} \cos \psi\,\sin\theta)=0\ , \\ 
	\label{eq:rev}    
	&{\rm Re} v = \frac{e^A}{2 \mu \sin\theta} d \theta \ ,\\
    \label{eq:dJpsi}
    &d\,\left(\frac1{\sin\theta} J_\psi \right) =2\mu e^{-A}{\rm Im} (v\wedge \omega_\psi)
\end{align}
\end{subequations}     
where again $J_\psi$ and $\omega_\psi$ are given by (\ref{eq:Jpsi}).

Then we have the fluxes, which are completely determined:
\begin{equation}
	H = d B   \, ,\qquad \mathbf{G} = e^{B \wedge} \mathbf{F}\,.
\end{equation}
So for example $G_2 = F_2 + B F_0$, but $G_0 = F_0$.  The $F_k$ are given by
\begin{subequations}
\begin{align}
    \label{eq:B}
 	B&= -\cot(\theta) J_\psi + \tan \psi \mathrm{Im}  \omega\ , \\
	F_0 &= -J_\psi\cdot d (e^{-\phi} \cos \psi {\rm Im} v)
	    + 5 \mu  e^{-A-\phi} \cos \psi\cos \theta \label{eq:F0}\ ,\\
	\label{eq:F2}
		F_2 &= F_0 \cot\theta J_\psi -J_\psi\cdot d\, {\rm Re} ( \cos \psi e^{-\phi}  v \wedge \omega_\psi) \\
		\nonumber &+ \mu \cos \psi  e^{-A-\phi}
		\left[(5+2\tan^2\psi) \sin\theta J_\psi +2 \sin\theta {\rm Re} v \wedge {\rm Im} v- 2\cos\theta\tan^2\psi {\rm Im} \omega_\psi\right]\ , \\
	F_4 & = F_0 \frac{J^2_\psi}{2 \sin^2\theta} + 
	d\Big[ \cos \psi \, e^{-\phi} (J_\psi \wedge {\rm Im} v - \cot\theta {\rm Re} (v \wedge\omega_\psi))\Big] \label{eq:F4}\ ,
	\\
	F_6 & = -\frac1{\cos^2\psi}{\rm vol}_6 \left(F_0\frac{\cos\theta}{\sin^3\theta} 
	+ 3 \frac{\mu \cos \psi e^{-\phi}}{\sin\theta}\right)\ .
\end{align}
\end{subequations} 
$F_4$ is automatically closed; this implies the Bianchi identity for $G_4$, which is $d G_4 + H \wedge G_2=0$.

\subsection{$SU(3)\times SU(3)$  $\theta = 0$} 
\label{sub:teq0}

The case $\theta=0$ has been discussed in subsection \ref{ssub:t0}. Here we will show that for this case, $G_6=\left( e^{b} \bf F\right)|_6= 0$, as claimed in the main text. 

Explicitly, what we have to compute is:
\begin{align}
G_6=\cancelto{0}{F_6}+\cancelto{0}{\frac{1}{3!}F_0b_\phi^3}+\frac{1}{2}F_2\wedge b_\phi\wedge b_\phi+F_4\wedge b_\phi\ ,
\label{g6}
\end{align}
where we are already using \eqref{su(2)prop} and \eqref{f6su3su3}. Taking into account the expression for $F_2$ -eq \eqref{eq:F2t0}-,  $F_4$ -eq.\eqref{eq:F2t0}- and $b=\tan(\psi)\im{\omega}$, \eqref{g6} reads: 
\begin{align}
\label{g6expand}
G_6&=\rho e^{-3A}\cos(\psi)^2\im\omega_\psi\wedge\re v\wedge \im v \wedge d\im v\nonumber \\&-\frac{1}{2}\tan^2(\psi)\im\omega\wedge\im\omega\wedge J_\psi^{-1} \llcorner d\re \left(  \rho e^{-3A} v \wedge \om_\psi\right) \, .
\end{align}
Let us massage the second term:
\begin{align}
&-\frac{1}{2}\tan^2(\psi)\im\omega\wedge\im\omega\wedge J_\psi^{-1} \llcorner d\re \left(  \rho e^{-3A} v \wedge \om_\psi\right)\nonumber\\ &=-\frac{1}{2}\tan^2(\psi)j^2\wedge J_\psi^{-1} \llcorner d\re \left(  \rho e^{-3A} v \wedge \om_\psi\right)\label{hola}\, ,
\end{align}
writing $j^2$ as $$j^2=\cos^2(\psi)J_\psi^2-\frac{2\cos^2(\psi)}{\tan^2(\psi)}j\wedge\re v\wedge\im v=\cos^2(\psi)^2J_\psi^2-\frac{2\cos^2(\psi)}{\tan^2(\psi)}J_\psi\wedge\re v\wedge \im v \, ,$$ to obtain:
\begin{align}
\label{tedioso}
-\frac{\sin^2\left(\psi^2\right)}{2} J_\psi^2\wedge J_\psi^{-1} \llcorner d\re \left(  \rho e^{-3A} v \wedge \om_\psi\right)+\cos^2\left(\psi^2\right)\re v\wedge\im v \wedge J_\psi \wedge J_\psi^{-1} \llcorner d\re \left(  \rho e^{-3A} v \wedge \om_\psi\right)\, .
\end{align}
We can use now -see \cite{Saracco:2012wc}-:
\begin{align}
&\label{comm}\left[J_\psi^{-1}\llcorner, J_\psi\wedge\right]=h\, ,\quad\quad 		h\omega_k\equiv (3-k)\omega_k\, ,
\end{align}
to rewrite \eqref{tedioso} as:
\begin{align}
&-\frac{\sin^2\left(\psi^2\right)}{2}J_\psi\wedge\left(J_\psi^{-1}\llcorner J_\psi\wedge+1\right)d\re \left(  \rho e^{-3A} v \wedge \om_\psi\right)+\nonumber\\&+\cos^2\left(\psi^2\right)\re v\wedge\im v\wedge\left(J_\psi^{-1}\llcorner J_\psi\wedge+1\right)d\re \left(  \rho e^{-3A} v \wedge \om_\psi\right)\, .
\label{aux2}
\end{align}
Finally, taking into account that the supersymmetry equations imply:
\begin{align}
J_\psi\wedge d\left(\rho e^{nA}\re (v\wedge\omega_\psi)\right)&=0\, ,   &  d\re v&=dA\wedge \re v\, ,
\end{align}
the second term of \eqref{g6expand} can be written as:
\begin{align}
 -\rho e^{-3A}\cos^2\left(\psi^2\right)\im\omega_\psi\wedge\re v\wedge\im v\wedge d\im v\, ,   
\end{align}
and therefore:
\begin{align}
    G_6=0\, .
\end{align}

\section{Proof of the source balanced equation}
\label{ap:SBEproof}

Let us show how the source balanced equation \eqref{preGOE} can be derived. First consider the following Mukai pairing
\be
\left<d_{H} \hat{F} , e^A \mathrm{Im\;}\Phi_- \right> = \left< \hat{F} , d_{H} \left(e^A \mathrm{Im\;}\Phi_- \right) \right> + dX_5 \;,
\ee
with $X_5$ defined as in \eqref{X5}. We can evaluate the left-hand side using the Bianchi identity \eqref{IIABI} in the presence of O6-planes and D6-branes, while the right-hand side can be evaluated using the supersymmetry equation \eqref{eq:psp-}. We obtain
\be
-3\mu  \left< \hat{F}, \mathrm{Im\;}\Phi_+ \right> + e^{4A} \left<\hat{F} , \star \lambda \left(\hat{F} \right) \right>  + dX_5 
= \left< \delta^{(3)}_{\mathrm{source}} , e^A  \mathrm{Im\;}\Phi_- \right> \;.
\label{pp1}
\ee
This expression can be rewritten by noting that taking the Mukai pairing of \eqref{eq:psp-} with $\Phi_+$ yields
\be
\mu \left< \mathrm{Re\;}\Phi_+ , \mathrm{Im\;}\Phi_+ \right> = e^{4A} \left<\Phi_+,  \star \lambda\left( \hat{F} \right) \right>\;.
\label{f03eq}
\ee
The existence of the $SU(3)\times SU(3)$-structure implies a generalised Hodge decomposition of the space of polyforms, according to their eigenvalues under two generalised complex structures $\left({\cal J}_+,{\cal J}_-\right)$. Under this decomposition $\Phi_+$ is of type $\left(3,0\right)$. This means that the right-hand side of (\ref{f03eq}) only receives a contribution from the $\left(-3,0\right)$ component of $\star \lambda\left( \hat{F} \right)$, and so we can replace $\star \lambda\left( \hat{F} \right)$ with $-i \hat{F}$ (see, for example \cite{Tomasiello:2007zq}). We can therefore write (\ref{f03eq}) as
\be
\mu \left< \mathrm{Re\;}\Phi_+ , \mathrm{Im\;}\Phi_+ \right> = -i e^{4A} \left< \Phi_+,\hat{F}  \right> \;.
\label{pppf}
\ee
Now using (\ref{pppf}) we have that (\ref{pp1}) reads
\be
3\mu^2 e^{-4A} \left< \mathrm{Re\;}\Phi_+ , \mathrm{Im\;}\Phi_+ \right>  - e^{4A} \sum_k \hat{F}_k \wedge \star \hat{F}_k + dX_5 
=  \left< \delta^{(3)}_{\mathrm{source}} , e^A  \mathrm{Im\;}\Phi_- \right> \, ,
 \label{ap:preGOE}
\ee
proving the desired relation.

It is important to note that integrating (\ref{ap:preGOE}) over the manifold leads to a constraint which does not differentiate between a local and smeared source, and therefore can be solved already for the $SU(3)$-structure case. If $X_5$ was a completely general function, then the solution to the integral of (\ref{preGOE}) would guarantee a local solution for some choice of $X_5$. However, $X_5$ is not an independent function, it is fixed by the fluxes and the polyforms, and therefore such a local solution is not guaranteed.

\bibliographystyle{jhep}
\bibliography{swampland}

\providecommand{\href}[2]{#2}\begingroup\raggedright\begin{thebibliography}{10}

\bibitem{Tsimpis:2012tu}
D.~Tsimpis, {\it {Supersymmetric AdS vacua and separation of scales}},  {\em
  JHEP} {\bf 08} (2012) 142, [\href{http://arxiv.org/abs/1206.5900}{{\tt
  arXiv:1206.5900}}].

\bibitem{Gautason:2015tig}
F.~F. Gautason, M.~Schillo, T.~Van~Riet, and M.~Williams, {\it {Remarks on
  scale separation in flux vacua}},  {\em JHEP} {\bf 03} (2016) 061,
  [\href{http://arxiv.org/abs/1512.00457}{{\tt arXiv:1512.00457}}].

\bibitem{Gautason:2018gln}
F.~F. Gautason, V.~Van~Hemelryck, and T.~Van~Riet, {\it {The Tension between
  10D Supergravity and dS Uplifts}},  {\em Fortsch. Phys.} {\bf 67} (2019),
  no.~1-2 1800091, [\href{http://arxiv.org/abs/1810.08518}{{\tt
  arXiv:1810.08518}}].

\bibitem{Blumenhagen:2019vgj}
R.~Blumenhagen, M.~Brinkmann, and A.~Makridou, {\it {Quantum Log-Corrections to
  Swampland Conjectures}},  {\em JHEP} {\bf 02} (2020) 064,
  [\href{http://arxiv.org/abs/1910.10185}{{\tt arXiv:1910.10185}}].
  [JHEP20,064(2020)].

\bibitem{Font:2019uva}
A.~Font, A.~Herr\'aez, and L.~E. Ib\'a\~nez, {\it {On scale separation in type
  II AdS flux vacua}},  {\em JHEP} {\bf 03} (2020) 013,
  [\href{http://arxiv.org/abs/1912.03317}{{\tt arXiv:1912.03317}}].

\bibitem{Apruzzi:2019ecr}
F.~Apruzzi, G.~Bruno De~Luca, A.~Gnecchi, G.~Lo~Monaco, and A.~Tomasiello, {\it
  {On AdS$_7$ stability}},  \href{http://arxiv.org/abs/1912.13491}{{\tt
  arXiv:1912.13491}}.

\bibitem{Vafa:2005ui}
C.~Vafa, {\it {The String landscape and the swampland}},
  \href{http://arxiv.org/abs/hep-th/0509212}{{\tt hep-th/0509212}}.

\bibitem{Brennan:2017rbf}
T.~D. Brennan, F.~Carta, and C.~Vafa, {\it {The String Landscape, the
  Swampland, and the Missing Corner}},  {\em PoS} {\bf TASI2017} (2017) 015,
  [\href{http://arxiv.org/abs/1711.00864}{{\tt arXiv:1711.00864}}].

\bibitem{Palti:2019pca}
E.~Palti, {\it {The Swampland: Introduction and Review}},  {\em Fortsch. Phys.}
  {\bf 67} (2019), no.~6 1900037, [\href{http://arxiv.org/abs/1903.06239}{{\tt
  arXiv:1903.06239}}].

\bibitem{Lust:2019zwm}
D.~Lust, E.~Palti, and C.~Vafa, {\it {AdS and the Swampland}},  {\em Phys.
  Lett.} {\bf B797} (2019) 134867, [\href{http://arxiv.org/abs/1906.05225}{{\tt
  arXiv:1906.05225}}].

\bibitem{DeWolfe:2005uu}
O.~DeWolfe, A.~Giryavets, S.~Kachru, and W.~Taylor, {\it {Type IIA moduli
  stabilization}},  {\em JHEP} {\bf 07} (2005) 066,
  [\href{http://arxiv.org/abs/hep-th/0505160}{{\tt hep-th/0505160}}].

\bibitem{Camara:2005dc}
P.~G. Camara, A.~Font, and L.~E. Ib{\'a}{\~n}ez, {\it {Fluxes, moduli fixing
  and MSSM-like vacua in a simple IIA orientifold}},  {\em JHEP} {\bf 09}
  (2005) 013, [\href{http://arxiv.org/abs/hep-th/0506066}{{\tt
  hep-th/0506066}}].

\bibitem{Grimm:2004ua}
T.~W. Grimm and J.~Louis, {\it {The Effective action of type IIA Calabi-Yau
  orientifolds}},  {\em Nucl. Phys.} {\bf B718} (2005) 153--202,
  [\href{http://arxiv.org/abs/hep-th/0412277}{{\tt hep-th/0412277}}].

\bibitem{Aharony:2008wz}
O.~Aharony, Y.~E. Antebi, and M.~Berkooz, {\it {On the Conformal Field Theory
  Duals of type IIA AdS(4) Flux Compactifications}},  {\em JHEP} {\bf 02}
  (2008) 093, [\href{http://arxiv.org/abs/0801.3326}{{\tt arXiv:0801.3326}}].

\bibitem{Acharya:2006ne}
B.~S. Acharya, F.~Benini, and R.~Valandro, {\it {Fixing moduli in exact type
  IIA flux vacua}},  {\em JHEP} {\bf 02} (2007) 018,
  [\href{http://arxiv.org/abs/hep-th/0607223}{{\tt hep-th/0607223}}].

\bibitem{Petrini:2013ika}
M.~Petrini, G.~Solard, and T.~Van~Riet, {\it {AdS vacua with scale separation
  from IIB supergravity}},  {\em JHEP} {\bf 1311} (2013) 010,
  [\href{http://arxiv.org/abs/1308.1265}{{\tt arXiv:1308.1265}}].

\bibitem{Saracco:2012wc}
F.~Saracco and A.~Tomasiello, {\it {Localized O6-plane solutions with Romans
  mass}},  {\em JHEP} {\bf 07} (2012) 077,
  [\href{http://arxiv.org/abs/1201.5378}{{\tt arXiv:1201.5378}}].

\bibitem{McOrist:2012yc}
J.~McOrist and S.~Sethi, {\it {M-theory and Type IIA Flux Compactifications}},
  {\em JHEP} {\bf 12} (2012) 122, [\href{http://arxiv.org/abs/1208.0261}{{\tt
  arXiv:1208.0261}}].

\bibitem{lin-lunin-maldacena}
H.~Lin, O.~Lunin, and J.~M. Maldacena, {\it {Bubbling AdS space and 1/2 BPS
  geometries}},  {\em JHEP} {\bf 0410} (2004) 025,
  [\href{http://arxiv.org/abs/hep-th/0409174}{{\tt hep-th/0409174}}].

\bibitem{afrt}
F.~Apruzzi, M.~Fazzi, D.~Rosa, and A.~Tomasiello, {\it {All AdS$_7$ solutions
  of type II supergravity}},  {\em JHEP} {\bf 1404} (2014) 064,
  [\href{http://arxiv.org/abs/1309.2949}{{\tt arXiv:1309.2949}}].

\bibitem{Couzens:2016iot}
C.~Couzens, {\it {Supersymmetric AdS$_{5}$ solutions of type IIB supergravity
  without D3 branes}},  {\em JHEP} {\bf 01} (2017) 041,
  [\href{http://arxiv.org/abs/1609.05039}{{\tt arXiv:1609.05039}}].

\bibitem{Junghans:2020acz}
D.~Junghans, {\it {O-plane Backreaction and Scale Separation in Type IIA Flux
  Vacua}},  \href{http://arxiv.org/abs/2003.06274}{{\tt arXiv:2003.06274}}.

\bibitem{Grana:2005sn}
M.~Gra\~na, R.~Minasian, M.~Petrini, and A.~Tomasiello, {\it {Generalized
  structures of N=1 vacua}},  {\em JHEP} {\bf 11} (2005) 020,
  [\href{http://arxiv.org/abs/hep-th/0505212}{{\tt hep-th/0505212}}].

\bibitem{Ibanez:2012zz}
L.~E. Ib{\'a}{\~n}ez and A.~M. Uranga, {\em {String theory and particle
  physics: An introduction to string phenomenology}}.
\newblock Cambridge University Press, 2012.

\bibitem{Bergshoeff:2001pv}
E.~Bergshoeff, R.~Kallosh, T.~Ortin, D.~Roest, and A.~Van~Proeyen, {\it {New
  formulations of D = 10 supersymmetry and D8 - O8 domain walls}},  {\em Class.
  Quant. Grav.} {\bf 18} (2001) 3359--3382,
  [\href{http://arxiv.org/abs/hep-th/0103233}{{\tt hep-th/0103233}}].

\bibitem{Marolf:2000cb}
D.~Marolf, {\it {Chern-Simons terms and the three notions of charge}},  in {\em
  {Quantization, gauge theory, and strings. Proceedings, International
  Conference dedicated to the memory of Professor Efim Fradkin, Moscow, Russia,
  June 5-10, 2000. Vol. 1+2}}, pp.~312--320, 2000.
\newblock \href{http://arxiv.org/abs/hep-th/0006117}{{\tt hep-th/0006117}}.

\bibitem{Taylor:1999ii}
T.~R. Taylor and C.~Vafa, {\it {R R flux on Calabi-Yau and partial
  supersymmetry breaking}},  {\em Phys. Lett.} {\bf B474} (2000) 130--137,
  [\href{http://arxiv.org/abs/hep-th/9912152}{{\tt hep-th/9912152}}].

\bibitem{Bielleman:2015ina}
S.~Bielleman, L.~E. Ib{\'a}{\~n}ez, and I.~Valenzuela, {\it {Minkowski 3-forms,
  Flux String Vacua, Axion Stability and Naturalness}},  {\em JHEP} {\bf 12}
  (2015) 119, [\href{http://arxiv.org/abs/1507.06793}{{\tt arXiv:1507.06793}}].

\bibitem{Carta:2016ynn}
F.~Carta, F.~Marchesano, W.~Staessens, and G.~Zoccarato, {\it {Open string
  multi-branched and K{\"a}hler potentials}},  {\em JHEP} {\bf 09} (2016) 062,
  [\href{http://arxiv.org/abs/1606.00508}{{\tt arXiv:1606.00508}}].

\bibitem{Herraez:2018vae}
A.~Herraez, L.~E. Ib{\'a}{\~n}ez, F.~Marchesano, and G.~Zoccarato, {\it {The
  Type IIA Flux Potential, 4-forms and Freed-Witten anomalies}},  {\em JHEP}
  {\bf 09} (2018) 018, [\href{http://arxiv.org/abs/1802.05771}{{\tt
  arXiv:1802.05771}}].

\bibitem{Marchesano:2019hfb}
F.~Marchesano and J.~Quirant, {\it {A Landscape of AdS Flux Vacua}},
  \href{http://arxiv.org/abs/1908.11386}{{\tt arXiv:1908.11386}}.

\bibitem{Palti:2008mg}
E.~Palti, G.~Tasinato, and J.~Ward, {\it {WEAKLY-coupled IIA Flux
  Compactifications}},  {\em JHEP} {\bf 06} (2008) 084,
  [\href{http://arxiv.org/abs/0804.1248}{{\tt arXiv:0804.1248}}].

\bibitem{Escobar:2018rna}
D.~Escobar, F.~Marchesano, and W.~Staessens, {\it {Type IIA flux vacua and
  $\alpha'$-corrections}},  {\em JHEP} {\bf 06} (2019) 129,
  [\href{http://arxiv.org/abs/1812.08735}{{\tt arXiv:1812.08735}}].

\bibitem{Escobar:2018tiu}
D.~Escobar, F.~Marchesano, and W.~Staessens, {\it {Type IIA Flux Vacua with
  Mobile D6-branes}},  {\em JHEP} {\bf 01} (2019) 096,
  [\href{http://arxiv.org/abs/1811.09282}{{\tt arXiv:1811.09282}}].

\bibitem{Tomasiello:2007zq}
A.~Tomasiello, {\it {Reformulating supersymmetry with a generalized Dolbeault
  operator}},  {\em JHEP} {\bf 02} (2008) 010,
  [\href{http://arxiv.org/abs/0704.2613}{{\tt arXiv:0704.2613}}].

\bibitem{Koerber:2010bx}
P.~Koerber, {\it {Lectures on Generalized Complex Geometry for Physicists}},
  {\em Fortsch. Phys.} {\bf 59} (2011) 169--242,
  [\href{http://arxiv.org/abs/1006.1536}{{\tt arXiv:1006.1536}}].

\bibitem{gualtieri}
M.~Gualtieri, {\it {Generalized complex geometry}},
  \href{http://arxiv.org/abs/math/0401221}{{\tt math/0401221}}. Ph.D. Thesis
  (Advisor: Nigel Hitchin).

\bibitem{Grana:2006kf}
M.~Gra\~na, R.~Minasian, M.~Petrini, and A.~Tomasiello, {\it {A Scan for new
  N=1 vacua on twisted tori}},  {\em JHEP} {\bf 05} (2007) 031,
  [\href{http://arxiv.org/abs/hep-th/0609124}{{\tt hep-th/0609124}}].

\bibitem{Behrndt:2004km}
K.~Behrndt and M.~Cvetic, {\it {General N = 1 supersymmetric flux vacua of
  (massive) type IIA string theory}},  {\em Phys. Rev. Lett.} {\bf 95} (2005)
  021601, [\href{http://arxiv.org/abs/hep-th/0403049}{{\tt hep-th/0403049}}].

\bibitem{Lust:2004ig}
D.~L{\"u}st and D.~Tsimpis, {\it {Supersymmetric AdS(4) compactifications of
  IIA supergravity}},  {\em JHEP} {\bf 02} (2005) 027,
  [\href{http://arxiv.org/abs/hep-th/0412250}{{\tt hep-th/0412250}}].

\bibitem{Koerber:2007jb}
P.~Koerber and L.~Martucci, {\it {D-branes on AdS flux compactifications}},
  {\em JHEP} {\bf 01} (2008) 047, [\href{http://arxiv.org/abs/0710.5530}{{\tt
  arXiv:0710.5530}}].

\bibitem{Gaiotto:2009mv}
D.~Gaiotto and A.~Tomasiello, {\it {The gauge dual of Romans mass}},  {\em
  JHEP} {\bf 01} (2010) 015, [\href{http://arxiv.org/abs/0901.0969}{{\tt
  arXiv:0901.0969}}].

\bibitem{Gibbons:1986df}
G.~W. Gibbons and N.~S. Manton, {\it {Classical and Quantum Dynamics of BPS
  Monopoles}},  {\em Nucl. Phys.} {\bf B274} (1986) 183--224.

\bibitem{Sen:1997kz}
A.~Sen, {\it {A Note on enhanced gauge symmetries in M and string theory}},
  {\em JHEP} {\bf 09} (1997) 001,
  [\href{http://arxiv.org/abs/hep-th/9707123}{{\tt hep-th/9707123}}].

\bibitem{Seiberg:1996nz}
N.~Seiberg and E.~Witten, {\it {Gauge dynamics and compactification to
  three-dimensions}},  in {\em {The mathematical beauty of physics: A memorial
  volume for Claude Itzykson. Proceedings, Conference, Saclay, France, June
  5-7, 1996}}, pp.~333--366, 1996.
\newblock \href{http://arxiv.org/abs/hep-th/9607163}{{\tt hep-th/9607163}}.

\bibitem{Hanany:2000fw}
A.~Hanany and B.~Pioline, {\it {(Anti-)instantons and the Atiyah-Hitchin
  manifold}},  {\em JHEP} {\bf 07} (2000) 001,
  [\href{http://arxiv.org/abs/hep-th/0005160}{{\tt hep-th/0005160}}].

\bibitem{Gaiotto:2009yz}
D.~Gaiotto and A.~Tomasiello, {\it {Perturbing gauge/gravity duals by a Romans
  mass}},  {\em J. Phys.} {\bf A42} (2009) 465205,
  [\href{http://arxiv.org/abs/0904.3959}{{\tt arXiv:0904.3959}}].

\bibitem{Hitchin:1999fh}
N.~J. Hitchin, {\it {Lectures on special Lagrangian submanifolds}},  {\em
  AMS/IP Stud. Adv. Math.} {\bf 23} (2001) 151--182,
  [\href{http://arxiv.org/abs/math/9907034}{{\tt math/9907034}}].

\bibitem{Hitchin:2000jd}
N.~J. Hitchin, {\it {The Geometry of Three-Forms in Six Dimensions}},  {\em J.
  Diff. Geom.} {\bf 55} (2000), no.~3 547--576,
  [\href{http://arxiv.org/abs/math/0010054}{{\tt math/0010054}}].

\bibitem{Koerber:2007hd}
P.~Koerber and D.~Tsimpis, {\it {Supersymmetric sources, integrability and
  generalized-structure compactifications}},  {\em JHEP} {\bf 08} (2007) 082,
  [\href{http://arxiv.org/abs/0706.1244}{{\tt arXiv:0706.1244}}].

\bibitem{Berkooz:1996km}
M.~Berkooz, M.~R. Douglas, and R.~G. Leigh, {\it {Branes intersecting at
  angles}},  {\em Nucl. Phys.} {\bf B480} (1996) 265--278,
  [\href{http://arxiv.org/abs/hep-th/9606139}{{\tt hep-th/9606139}}].

\bibitem{Shandera:2003gx}
S.~Shandera, B.~Shlaer, H.~Stoica, and S.~Tye, {\it {Interbrane interactions in
  compact spaces and brane inflation}},  {\em JCAP} {\bf 02} (2004) 013,
  [\href{http://arxiv.org/abs/hep-th/0311207}{{\tt hep-th/0311207}}].

\bibitem{Andriot:2019hay}
D.~Andriot and D.~Tsimpis, {\it {Gravitational waves in warped
  compactifications}},  \href{http://arxiv.org/abs/1911.01444}{{\tt
  arXiv:1911.01444}}.

\bibitem{Aharony:2010af}
O.~Aharony, D.~Jafferis, A.~Tomasiello, and A.~Zaffaroni, {\it {Massive type
  IIA string theory cannot be strongly coupled}},  {\em JHEP} {\bf 11} (2010)
  047, [\href{http://arxiv.org/abs/1007.2451}{{\tt arXiv:1007.2451}}].

\bibitem{Buratti:2020kda}
G.~Buratti, J.~Calderon, A.~Mininno, and A.~M. Uranga, {\it {Discrete
  Symmetries, Weak Coupling Conjecture and Scale Separation in AdS Vacua}},
  \href{http://arxiv.org/abs/2003.09740}{{\tt arXiv:2003.09740}}.

\end{thebibliography}\endgroup

\end{document}